\documentclass[11pt,a4paper]{article}

\usepackage[utf8]{inputenc} 
\usepackage[T1]{fontenc}    %
\usepackage[english]{babel} 
\usepackage[final]{microtype} 

\usepackage{amssymb,amsmath,mathtools} 
\usepackage{amsthm} 

\usepackage{xcolor}
\usepackage{bbm}
\usepackage{braket}
\usepackage{hyperref}
\usepackage[most]{tcolorbox}

\usepackage[hmargin=0.12\paperwidth,vmargin=0.16\paperwidth,bindingoffset=0cm]%
{geometry} 

\numberwithin{equation}{section} 

\theoremstyle{plain}
  \newtheorem{thm}{Theorem}
  \newtheorem{prop}{Proposition}
  \newtheorem{lemma}{Lemma}
  \newtheorem{cor}{Corollary}
\theoremstyle{definition}
  \newtheorem{definition}{Definition}
  \newtheorem{remark}{Remark}
  
  \numberwithin{prop}{section}
   \numberwithin{cor}{section}
   \numberwithin{remark}{section}

 \newcommand{\C}{\mathbb{C}}

\DeclareMathOperator{\erfc}{erfc} 

\newcommand{\rbox}[1]{\begin{tcolorbox}[arc=0mm,oversize,colback=purple!3!white,colframe=purple!100!white]#1\end{tcolorbox}}

\newcommand{\Eq}[2]{\begin{equation}\label{#1}\begin{aligned}#2 \end{aligned}\end{equation}}

\newcommand{\theo}[2]{\rbox{\begin{thm}\label{#1} #2 \end{thm}}}
\newcommand{\coro}[2]{\rbox{\begin{cor}\label{#1} #2 \end{cor}}}

\newcommand{\propo}[2]{\rbox{\begin{prop}\label{#1} #2 \end{prop}}}
\newcommand{\rem}[2]{\begin{remark}\label{#1} #2 \end{remark}}

\newcommand{\tp}{\mathcal{T}_{1,b}^+}
\newcommand{\tm}{\mathcal{T}_{1,b}^-}
\newcommand{\lrbra}[1]{\left( #1 \right)}
\newcommand{\facb}[1]{\frac{1}{b}\sqrt{\frac{2 #1}{\Delta q(b)}}}

\title{\Large\bfseries Large $N$ expansions of the partition function for Coulomb systems on the surface of a cylinder}%
   
\author{Bo-Jian Shen and Peter J. Forrester}
\date{}

\begin{document}

\maketitle

School of Mathematics and Statistics,  The University of Melbourne,
Victoria 3010, Australia. \: \: Email: 
{\tt bojian.shen@unimelb.edu.au};
{\tt pjforr@unimelb.edu.au}   \\

\begin{abstract}
The two-dimensional one-component plasma forms a droplet in the case that the one-body potential is proportional to the number of charges. Our interest is in studying the large $N$ form of the corresponding partition function, with the plasma confined to the surface of the cylinder. The one-body potential is (up to technical restrictions) allowed to be arbitrary in the direction of the axis of the cylinder, whereas the coupling is restricted to the exactly solvable value $\beta = 2$. It is demonstrated that a change of variables can be made to an annular droplet plasma model, up to a certain Jacobian factor. The latter can be interpreted as the characteristic function of a certain linear statistic, which generalises the original aim of our study to also analysing the large $N$ form of the characteristic function for the annular droplet plasma model. While this is well known in the case of soft wall boundary conditions, our analysis (based on  Laplace's method applied to certain integrals, and the Euler-Maclaurin summation formula) covers the case of one or two hard walls at the boundary, or inside, of the droplet for which the soft edge formulas are shown to no longer hold. Use of this to study the large $N$ expansion of the cylinder partition function requires knowledge of the same expansion in the annular case. This is not available in the case of two hard walls inside the droplet, which we treat instead via a direct analysis. In distinction to the case of the annular partition function, the large $N$ expansion of the logarithm of the cylinder partition
function is found to always have no term proportional to $\log N$, independent of the presence of hard walls.
\end{abstract}

\section{Introduction}
\subsection{The 2D OCP and its partition function for large $N$ --- disk and cylinder geometry}
The two-dimensional one-component plasma (2D OCP), or Coulomb gas, is a basic statistical physics model for $N$ mobile charges of the same sign confined to a two dimensional domain $\Omega$ and neutralised by an opposite background charge of density $\rho_{\rm b}$ \cite{Fo10}. The neutrality condition fixes the total background charge to match the total particle charge. In general, the background, together with the imposed boundary condition, determines the region where the particles concentrate in equilibrium at absolute temperature $T$, which is also known as the droplet. The probability density function for the event that the particles are at positions $\vec{r}_{1}, \ldots, \vec{r}_{N}$ is given as
$$
\frac{1}{\hat{Z}_{N}} e^{-\beta U\left(\vec{r}_{1}, \ldots, \vec{r}_{N}\right)},
$$
where $U\left(\vec{r}_{1}, \ldots, \vec{r}_{N}\right)$ is the total potential energy of the system, $\beta$ is the (dimensionless in our units) inverse temperature, and the normalization $\hat{Z}_{N}$ is given by
\begin{equation}\label{0.0a}
\hat{Z}_{N}=\int_{\Omega} d \vec{r}_{1} \cdots \int_{\Omega} d \vec{r}_{N} e^{-\beta U\left(\vec{r}_{1}, \ldots, \vec{r}_{N}\right)}.
\end{equation}
The term $e^{-\beta U\left(\vec{r}_{1}, \ldots, \vec{r}_{N}\right)}$ is referred to as the Boltzmann factor and $\hat{Z}_{N} / N!=: Z_{N}$ is called the (canonical) partition function.
The potential $U\left(\vec{r}_{1}, \ldots, \vec{r}_{N}\right)$ is obtained from the electrostatic energy consisting of particle-particle repulsion, particle-background attraction and the background-background interaction. In the two-dimensional planar case, the Coulomb interaction is logarithmic: the particles at positions $\vec{r}$ and $\vec{r}^{\,\prime}$ interact via potential $\Phi(\vec r,\vec r^{\,\prime}):=-\log |\vec r-\vec r^{\,\prime}|$, while on a general surface or a domain with boundary, $\Phi(\vec r,\vec r^{\,\prime})$ is replaced by the appropriate Green's function solving the Poisson equation on the surface with the prescribed boundary condition; see Appendix A for the example of the cylinder.
The 2D OCP on a cylinder domain is to be the focus of our
study; for earlier works on this model see
\cite{Ch81,CFS83,SWK04,CFTW15,Fa19,CSA25}.

The feature of our interest is
the large $N$ expansion of the logarithm of the partition function.
It is appropriate to revise what is already known on this from earlier work. From the calculation in Appendix A of the 
pair potential on the surface of the cylinder, we know that the latter can be viewed as the rectangle
$x \in [0,W]$, $y \in [-L/2,L/2]$ contained in $\mathbb R^2$ with periodic boundary conditions of period $W$ in the $x$ direction.  Here $W$ is the circumference length of the cylinder, and $L$ is the height of the cylinder. After the calculation of the total potential energy in this setting, let $\Omega$ in (\ref{0.0a}) be replaced by $\mathbb R^2$ --- this is the setting of a so-called soft wall. 
It is predicted in \cite{Fo91}, \cite{CFTW15} that for this with
 $N$ large, $\rho_{\rm b}$ fixed,
\begin{equation}\label{0.11}
 \beta F =  - \log Z_N = N \Big ( 1 - {\beta \over 4} \Big ) \log \rho_{\rm b} + N \beta f(\beta) + 2 W \sqrt{\rho_{\rm b}}
  \bigg ( {2 \log (\beta / 2) \over 3 \pi} \bigg ) + 
 {\pi L \over 6 W} + 
 {\rm O}(1).
 \end{equation}
 The case $\beta = 2$ is exactly solvable due to an analogy with the
 absolute value squared of a certain two-dimensional free Fermi 
 many body wave function --- that holds true too for a planar
 domain \cite{AJ81}, the sphere \cite{Ca81}, the topological torus \cite{Fo06}, 
 pseudosphere \cite{JT98},
and Flamm's paraboloid \cite{FT08}, \cite{Fa19}
 --- and it is found
 that (\ref{0.11}) holds with $\beta f (\beta) |_{\beta = 2} =
 - {1 \over 2} \log(2 \pi^2)$, and with all correction terms identically zero \cite{CFTW15}. In \cite{CFTW15}, numerical evidence relating to $\beta = 4,6$ and 8 is given that the unspecified ${\rm O}(1)$ term in (\ref{0.11})
 is independent of the dimensionless parameter
 $\rho_{\rm b} W^2$, or equivalently independent of $L/W$.
 The term $\pi L/(6W)$ is shown in \cite{Fo91} to arise
 as a consequence of the relationship of the $L \to \infty$
 limit and the corresponding Gaussian free field.

It is instructive to compare (\ref{0.11}) to the corresponding expansion for a disk geometry. Taking the radius to be $R$, and allowing for soft walls by replacing
$\Omega$ in (\ref{0.0a}) with $\mathbb R^2$ as assumed for
(\ref{0.11}), one has the prediction \cite{JMP94},
\cite{CFTW15} that for
large $N$ with $\rho_{\rm b}$ fixed
\begin{equation}\label{0.11a}
\beta F =  - \log Z_N = N \Big ( 1 - {\beta \over 4} \Big ) \log \rho_{\rm b} + N \beta f(\beta) + 2 \pi R \sqrt{\rho_{\rm b}}
  \bigg ( {2 \log (\beta / 2) \over 3 \pi} \bigg ) + 
 {1 \over 12} \log N + 
 {\rm O}(1).
\end{equation}
Thus the leading terms are the same, and the first corrections agree too if one interprets $2W$ in (\ref{0.11}) and $2 \pi R$ in (\ref{0.11a}) as the length of the boundaries --- thus the first corrections have the interpretation as a surface tension term. However after this, the expansions then differ, with in particular there being no term proportional to $\log N$ in
(\ref{0.11}). This is in keeping with the prediction from
\cite{JMP94} that for a general two-dimensional domain, the free energy of the 2D OCP contains a single term at order $\log N$, which is explicitly
${\chi \over 12} \log N$. Here
$\chi$ is the Euler characteristic of $\Omega$ (e.g.~$\chi = 1$ for a disk, $\chi = 2$ for a sphere, $\chi=0$ for an annulus, cylinder or topological torus). For the exactly solvable coupling $\beta = 2$, this is verified in
\cite{JMP94} for the disk and sphere using product forms for
$Z_N$ obtained earlier in \cite{AJ81} (disk), 
\cite{Ca81} (sphere). Its validity at this coupling for the
topological torus is a feature of the exact result of \cite{Fo06}, and that for the annulus of exact results the large $N$ expansions obtained
in \cite{FF11}, \cite{BKS23}, \cite{BKSY25}. 
Higher genus Riemann surfaces are considered in
\cite{KMMW17,BKSY25,Bo25,SY25}. Numerical solutions are obtained for the couplings $\beta=4,6$ in \cite{TF99}.
In the case of the
soft edge disk and sphere, the large $N$ expansion of the first order term in an expansion of $\beta F$ in powers of $(\beta - 2)$ is also possible \cite{Sh11}, \cite{CFTW15}, \cite[Remark 4.1.1]{BF25} and consistency with the predicted asymptotic forms (\ref{0.11}) and (\ref{0.11a}) is obtained.

The expansions (\ref{0.11}) and (\ref{0.11a}) are valid under the assumption of soft wall boundary conditions. The work \cite{JT96} has considered the modification of 
(\ref{0.11a})
due to Dirichlet boundary conditions. 
While the surface tension term changes, it is
 predicted that the universal ${1 \over 12} \log N$ term remains unchanged. This same scenario is predicted for hard wall boundary conditions at the extremity of the background, with it being shown in \cite{JMP94} for the coupling $\beta = 2$ that the term proportional to $R$ in the expansion
 (\ref{0.11a}) is now equal to $2 \pi R \gamma_{\partial \Omega}$ with
 \begin{equation}\label{0.11b}
 \gamma_{\partial \Omega} = - \Big ( \frac{\rho_{\rm b} }{ 2 \pi} \Big )^{1/2} \int_0^\infty \log
\Big ( \frac{1 }{ 2} \left(1 + {\rm erf} (y)\right) \Big ) \, dy.
\end{equation} 
This was first identified in \cite{FS82}. Missing from the literature is the analogous calculation for cylinder geometry in the case of hard wall boundary conditions
at either of both boundaries of support. 
It follows from our Proposition \ref{P1.1} below
that for $\beta =2$ the term proportional to $W$ in (\ref{0.11}) is to be replaced by $2W \gamma_{\partial \Omega}$ (two hard walls) or $W \gamma_{\partial \Omega}$
(one hard wall), where
$ \gamma_{\partial \Omega}$ is given by (\ref{0.11b}).

\subsection{Droplet reformulation of the 2D OCP --- annular geometry}

We are also interested in the large $N$ expansion of the logarithm of the partition function for a
reformulation of the 2D OCP model obtained by taking the background density as proportional to $N$. The limiting eigenvalue support is then in general compact, and referred to as the droplet. Moreover, the potential energy between a charge and the background is then of the form $N Q(\vec{r})/ 2 + K_N$ for some $Q(\vec{r})$ independent of $N$, $K_N$ 
independent of $\vec{r}$. This draws attention to the functional form
\begin{align}\label{1.9}
    e^{-N \beta\sum_{j=1}^N Q(z_j)/2} 
 \prod_{1 \le j < k \le N} | z_k - z_j|^{\beta}.
\end{align}
In the case that $Q(z) = Q(|z|)$
we will refer to this as the annular droplet Boltzmann factor.
When $\beta=2$, up to normalisation (\ref{1.9}) is equal to the eigenvalue probability density function for the class of complex normal random  matrices $M M^\dagger = M^\dagger M$ distributed according to the matrix probability density proportional to $e^{- N {\rm Tr} \, Q(M)}$; see e.g.~\cite[\S 5]{BF25}.
The normalisation for (\ref{1.9}) is the configuration integral
\begin{align}\label{1.9a}
    \hat{Z}_{N,\beta}(Q) := 
 \int_{ \Omega} d^2 z_1 \cdots \int_{\Omega} d^2 z_N \,
 e^{-\beta N \sum_{j=1}^N Q(z_j)/2} \prod_{1 \le j < k \le N} |z_k - z_j|^\beta.
\end{align}

Analogous to the question of the large $N$ expansion of the free energy for the 2D OCP, in relation to (\ref{1.9}), one seeks the large $N$ expansion of ${Z}_{N,\beta}(Q) :=
\frac{1 }{ N!} \hat{Z}_{N,\beta}(Q)$
as initiated in \cite{WZ03,ZW06}. On this, one knows from
\cite[Corollary 1.1]{LS17} that
\begin{equation}\label{eq:ZW-conj}
   \log {Z}_{N,\beta}(Q) 
 = - {\beta \over 2}  N^2 I_Q[\mu_Q] - 
  \Big ( 1 - {\beta \over 4} \Big )
 N\log N + \Big( C(\beta) +
 \Big ( 1 - {\beta \over 4} \Big ) E_Q[\mu_Q]\Big)N + \cdots,
\end{equation}
where
$$
I_Q[\mu]:=\iint_{\C^2}\log\frac{1}{|z-w|}\,d\mu(z)\,d\mu(w)+\int_{\C}Q\,d\mu,
\qquad
E_Q[\mu_Q]:=\int_{\C}\log\left(\frac{\Delta Q}{4}\right)\,d\mu_Q.
$$
Here $I_Q[\mu], E_Q[\mu_Q]$
are respectively the weighted logarithmic energy of $\mu$ (minimised over compactly supported probability measures by the equilibrium measure $\mu_Q$) and the entropy of $\mu_Q$ --- here $\Delta$ denotes the usual Laplacian operator, so that 
\begin{equation}\label{1.7a}
d\mu_Q=\tfrac{1}{4}\Delta Q\,dA
\end{equation}
on the droplet in terms of the normalized area measure $dA(z)={d^2z}/{\pi}$. 
In the case of a constant background at least, 
the lower order terms
in (\ref{eq:ZW-conj}) are expected to be  simply related to those in the corresponding fixed $\rho_{\rm b}$ version of
the model.

Set $\beta = 2$, $Q(z) = Q(|z|)$, and assume that $Q$ is subharmonic in $\mathbb C$, and strictly subharmonic in a neighbourhood of the droplet. Then the eigenvalue support is the annulus or disk
$r_1 > 1 > r_0 \ge  0$. Our attention is to be focussed on the case of the annulus $r_0 > 0$, and we write
${Z}_{N,\beta}(Q) = Z_N^{\rm a}(Q)$ where the superscript ``a'' is for annulus. In this setting,
more can be said in relation to higher order terms
in (\ref{eq:ZW-conj}), including in circumstances with a varying background \cite{BKS23}, and a (outer) hard wall 
\cite{AFLS25}. In the 2D OCP model of the previous subsection with disk geometry, the effect of a varying radial background and hard wall from the viewpoint of correlation functions was studied in the early work
\cite{AL84}.

Consider first the case of soft wall boundary conditions.
We have from \cite[Eq.~(1.16)]{BKS23} that the leading order term not shown in (\ref{eq:ZW-conj}) is the constant
\begin{equation}\label{F1}
 F_{\mathbb{S}\cap \mathbb{D}} [Q]:=\frac{1}{12}\log \Big (\frac{r_0^2\Delta Q(r_0)}{r_1^2\Delta Q(r_1)} \Big )-\frac{1}{16}\left(r_1\frac{\partial_r \Delta Q(r_1)}{\Delta Q(r_1)}-r_0\frac{\partial_r \Delta Q(r_0)}{\Delta Q(r_0)}\right)+\frac{1}{24}\int_{r_0}^{r_1}\left(\frac{\partial_r \Delta Q(r)}{\Delta Q(r)}\right)^2rdr
\end{equation}
Now impose a hard wall at $r=1$, which then becomes the outer boundary of the support. As is well known (see e.g.~\cite[\S 5.1] {BF25a}), the equilibrium measure acquires a singular component at the hard wall, which is a uniform measure containing all the mass  of the soft-wall equilibrium measure outside the hard wall (see \eqref{muq-ome}), also known as a balayage measure \cite{Ch23,AR17}. In this case with $r_0=0$ (support is a disk), explicit calculation for $\beta = 2$ carried out in  \cite{AFLS25} shows that with this coupling the coefficient of the $\log N$ term in \eqref{eq:ZW-conj} is modified from the prefactor
${1 \over 12}$ to the prefactor ${1 \over 3}$, and an additional $\sqrt{N}$ boundary contribution appears in the partition function (for $\beta = 2$ the term at this order in
\eqref{eq:ZW-conj} vanishes). Specifically, the large $N$ expansion of $\log Z_{N,2}(Q)$ acquires
\begin{align}\label{eq:intro-gamma-in}
-\sqrt{N\,\Delta Q(1)}\,(\gamma^{\rm in}+\gamma^{\rm out}),
\end{align}
where
\begin{align}\label{gi}
    \gamma^{\rm in} \;:=\; -\frac{1}{\sqrt{2}}\int_{-\infty}^{0}\log\!\Big(\tfrac{1}{2}\operatorname{erfc}(x)\Big)\,dx,\quad
\gamma^{\rm out}:=-\frac{1}{\sqrt{2}}\int_0^{\infty} \log\left( \sqrt{\pi} x e^{x^2} \erfc(x) \right) dx
\end{align}
are universal constants. The same integrals are also found earlier in the large gap asymptotics on annuli in the Ginibre ensemble \cite{Ch24,Fo92}, and in the complex and symplectic spherical ensembles with point charges \cite{BP26}. For related hard edge universality results on the local correlation kernel and on disk counting cumulants near a hard wall, see \cite{Se22,ACCL23}. Apart from the multiplicative factor, the integral in $\gamma^{\rm in}$ is identical to $\gamma_{\partial \Omega}$
\eqref{0.11b} after a change of variable $x\to-x$. Specifically,
\begin{align}\label{1.10a}
\gamma_{\partial \Omega} = \Big ( {\rho_{\rm b} \over \pi}
\Big )^{1/2} \gamma^{\rm in}.
\end{align}

In planar geometry with $\beta = 2$, other studies relating to the large $N$ expansion of the logarithm of the partition function have been carried out.
Thus the case of a non-radial doubly connected droplet 
is considered in \cite{BSY24,BYY26}; that of a non-radial
multi-component droplet in \cite{By25,BYY26};
the plasma with macroscopic holes in \cite{Ro25}; 
the case of a non-constant radial potential with or without
hard walls in \cite{BKS23,Ch24,AFLS25,No25,BCMS25}; the case with a hard wall along the boundary of a Jordan domain with corners and no confining potential in \cite{JV26};
the effect of the density vanishing on a circle inside the droplet, and of a spectral gap in \cite{AL25,ACC26}.
Closely related to the case of a hard wall with no confining potential is that of confinement to a contour. 
For $\beta =2$ and a smooth contour, a study of the asymptotics of the partition function has been undertaken in \cite{Jo22}. This was later extended to
general $\beta > 0$ in 
\cite{CJ25,WZ22}.

\subsection{Droplet reformulation of the 2D OCP --- cylinder geometry}

We would like to give a droplet reformulation of the 2D OCP on a cylinder. Provided the potential is constant in the periodic direction, the corresponding partition function can be computed exactly for the coupling $\beta = 2$, and its large $N$ expansion analysed. We will see that this can be done both for soft walls, and in the presence of hard walls, thus giving cylinder analogues of the planar
results of \cite{BKS23}, \cite{AFLS25}.
Moreover, due to a mapping from the cylinder  to the planar partition functions (see (\ref{eq5a}) below), we will see that generalising our problem is the question of the large $N$ asymptotic expansion of the characteristic function of a linear statistic in the annular case. While this is well known in the case of soft walls, it would seem that the existing literature does not extend to the hard wall case. Thus, as part of the present paper we undertake a study of this question, albeit with the restriction that the one-body potential is rotationally symmetric.

For the droplet reformulation of the 2D OCP on a cylinder 
at coupling $\beta =2$, we take for the analogue of
(\ref{1.9})
\begin{equation}\label{eq4}
(2 \pi )^{N} e^{-2 \pi N \sum_{l=1}^N (Q^{\rm c}(Y_l) + Y_l) } e^{2 \pi \sum_{l=1}^N Y_l}
\prod_{1 \le j < k \le N} \Big |  e^{2 \pi i (X_j-i Y_j)} - e^{2 \pi i (X_k -i Y_k)}  \Big |^2, \quad 0\le X_k \le 1,
\end{equation}
where the one-body potential $Q^{\rm c}$ is independent of
the periodic direction.
See (\ref{eq1B}) below and associated text for justification.
We will refer to this as the cylinder droplet Boltzmann factor.
Denote the corresponding normalisation by
$ \hat{Z}_N^{\rm c}(Q^{\rm c})$, and partition function ${Z}_N^{\rm c}(Q^{\rm c}) = {1 \over N!} \hat{Z}_N^{\rm c}(Q^{\rm c})$.
By a simple change of variables detailed in
\S \ref{S2.2.1}, we will see that 
this cylinder partition function can be mapped to the partition function for
the Boltzmann factor
\begin{equation}\label{eq5a}
\prod_{l=1}^N e^{-N  q(|z_l|)} |z_l|^{-1}  
 \prod_{1 \le j < k \le N}  |  z_j - z_k    |^2,
\end{equation}
where
\begin{equation}\label{eq5}
q(r) = 2 \pi Q^{\rm c} ((1/2 \pi) \log r  ) + \log r.
\end{equation}
We will refer to the partition function for
(\ref{eq5a}) as ${Z}_N^{\rm c}(q)$.
It will be assumed throughout that $q(r)$ is such that the
corresponding droplet is an annulus, denoted by $\mathbb{S}=\left\{z \in \mathbb{C}: r_0 \leq|z| \leq r_1\right\}$. One recalls from
e.g.~\cite{BKS23} that the conditions for the inner radius $r_0$ and outer radius $r_1$ are
\begin{equation}\label{rr01}
q'(r_0)=0, \qquad r_1 q'(r_1) = 2
\end{equation}
respectively.


From the change of variables linking (\ref{eq4}) and
(\ref{eq5a}) (this is further detailed in 
\S \ref{S2.2.1} below) we have
\begin{equation}\label{b.1}
\log Z_N^{\rm c}(Q^{\rm c}) =  N \log {1 \over 2 \pi} + 
\log Z_N^{\rm c}(q) =  N \log {1 \over 2 \pi} + 
\log Z_N^{\rm a}(q) + \log { Z_N^{\rm c}(q) \over
 Z_N^{\rm a}(q)}.
\end{equation}
Here $Z_N^{\rm c}(Q^{\rm c})$ is the partition function for
(\ref{eq4}), $ Z_N^{\rm a}(q)$ is the partition function for (\ref{1.9}) with $Q(z) = q(|z|)$ and $\beta = 2$, and
$ Z_N^{\rm c}(q)$ is the partition function for (\ref{eq5a}). The function $q(r)$ is related to $Q^{\rm c}$ by (\ref{eq5}). Following this line, our task is to determine the asymptotic expansion for each term on the right hand side of \eqref{b.1} in the presence of hard walls.

Denote by $\langle \cdot \rangle_{\rm a}$ the average with respect to the annular droplet Boltzmann factor. By definition, the logarithmic ratio in \eqref{b.1} is equivalent to
\begin{equation}\label{eq9}
\log
 \Big \langle e^{ - \sum_{l=1}^N \log |z_l|} \Big \rangle_{\rm a},
\end{equation}
which suggests a possible application of the Gaussian fluctuation formula. Gaussian fluctuations of smooth linear statistics are by now well understood for regular soft-edge planar Coulomb gases \cite{Fo99,RV07,AHM11,AHM15,Se23,BF23}, although variations on the theme (particularly when the linear statistic is not smooth) are still attracting present day studies \cite{LS25,MMO26,CGX26}. In the determinantal case, the rotationally invariant formula goes back to \cite{Fo99}; for the Ginibre ensemble and for more general random normal matrix potentials, the asymptotic distribution of the fluctuation about the equilibrium for non-radial test functions was established in \cite{RV07,AHM11,AHM15}. 
For a radially symmetric smooth test function $a(z)\in\mathcal{C}_0^{\infty}(\mathbb{C})$ and an annulus droplet with soft wall boundary conditions, \cite[specialisation of Theorem~7.4.1]{AHM11} gives that
\begin{align}\label{1.19}
    \log\left\langle e^{\sum_{i=1}^{N}a(|z_i|)}\right\rangle_{\mathrm a}=
N\int a\,d\mu_q
+M[a]
+\frac{1}{4}\int_{r_0}^{r_1}r(a'(r))^2dr
+{\rm o}(1),
\end{align}
where $\mu_q$ is the equilibrium measure (\ref{1.7a}) and
\begin{align}
    M[a]
=
\frac14\bigl[r a'(r)\bigr]\Big|_{r_0}^{r_1}
-
\frac14\int_{r_0}^{r_1}
\,a'(r)\frac{r\partial_r\Delta q(r)}{\Delta q(r)}\,dr.
\end{align}
Taking $a(z)=-\log |z|$ and simplifying shows that for large $N$
\begin{align}\label{eq9A}
\log  {Z_N^{\rm c}(q) \over Z_N^{\rm a}(q)}  = 
-\frac{N}{2}\int_{r_0}^{r_1} (\log r)
r\Delta q(r) \, dr  + {1 \over 4}
\log \frac{r_1\Delta q(r_1)}{r_0\Delta q(r_0)}+{\rm o}(1).
\end{align}
Specialising now to $q(r)$  in
(\ref{eq6}) below as corresponds to $Q^{\rm c}(Y) = {W \over L} Y^2$, one finds that the first integral vanishes, and the remaining term gives $-\pi L / 2 W$.

In this case, a direct approach is also possible.
Thus, for 
$q(r)$ specified in
(\ref{eq6}), the use of the exponential coordinates in
(\ref{eq4A}) and the formulas (\ref{logratioZ}), (\ref{eq10A}) allow us to compute
\begin{equation}\label{eq5X}
{Z_N^{\rm c}(q) \over Z_N^{\rm a}(q)} =
\prod_{l=0}^{N-1} {e^{\pi (N-2l-1)^2/(2 \rho_{\rm b} W^2) }  \over e^{\pi (N-2l-2)^2/(2 \rho_{\rm b} W^2) } }
= e^{-\pi L / 2 W},
\end{equation}
confirming the validity of the use of the fluctuation formula (and showing too that all correction terms vanish).
However, we cannot expect this ``quick'' approach to hold beyond the case of soft wall boundary conditions, as we would expect that (\ref{1.19}) requires modification in the presence of a hard wall or walls.

The details of the modification present themselves as a well motivated research question, which seems to have escaped attention in the extensive early literature as cited above.
We have the tools 
to investigate the hard wall effect for the fluctuation formula in the case that the function $a(z)$ is rotationally invariant and smooth in a neighbourhood of the droplet. The asymptotic expansion is then performed for the averaged linear statistic $\log\left< e^{ \sum_{i}a(\lvert z_{i} \rvert ) } \right>_{\mathrm{a}}$.
As expected, the placement of hard wall(s) inside the droplet alters the form of (\ref{eq9A}).
The precise formulas are documented in Theorem \ref{thm1}.

\theo{thm1}{
Let $[\rho_-,\rho_+]\subset (0,\infty)$ 
be the hard wall constraint. Denote by $\mu_q^{\Omega}$ the (hard-wall) equilibrium measure given by \eqref{muq-ome} below and write 
\begin{align}\label{ome+-}
    \omega_-:=\min\!\{\rho_+,\max\{\rho_-,r_0\}\},\qquad
    \omega_+:=\min\!\{\rho_+,\max\{\rho_-,r_1\}\}
\end{align}
for the endpoints of its support $I:=[\omega_-,\omega_+]$ (If $\omega_-=\omega_+=:\omega$, the support collapses to the circle $|z|=\omega$ and $d\mu_q^{\Omega}=d\sigma_\omega$).
As $N\to \infty$, we have the asymptotic expansion
\begin{align}\label{ea-asy-thm}
    \;\log\left\langle e^{\sum_{i}a(|z_i|)}\right\rangle_{\mathrm a}
	= N\int a\,d\mu^{\Omega}_q \;+\; \mathcal{E}_-[a] + \mathcal{E}_+[a]
	\;+\; \frac14\int_{I}r\,a'(r)\left(a'(r) - \frac{\partial_r\Delta q(r)}{\Delta q(r)}\right)dr + \mathrm{o}(1).
\end{align}
Here each of the boundary terms $\mathcal{E}_+,
\mathcal{E}_-$ depend on the nature of the upper $(+)$ and lower $(-)$ endpoints. 
\begin{itemize}
    \item For the overlapping case $([\rho_-,\rho_+]\cap (r_0,r_1)\neq \emptyset$ and $\omega_-<\omega_+)$, these are given by
\begin{align}\label{E-pm}
    \mathcal{E}_{\pm}[a]=
\begin{cases}
    \pm \frac{\omega_{\pm}a'(\omega_{\pm})}{4},\quad \text{soft edge},\\
    \mp \frac{\omega_{\pm}a'(\omega_{\pm})}{4}\Bigl(\log N + 2\log\dfrac{|m_{\pm}|}{\omega_{\pm}} + \log\dfrac{8\pi}{\Delta q(\omega_{\pm})} - 1\Bigr),\quad \text{hard wall strictly inside},\\
    \mp\frac{\omega_{\pm}a'(\omega_{\pm})}{4}\bigl(2\log 2-1\bigr),\quad \text{hard wall at the droplet boundary},
\end{cases}
\end{align}
where $m_{+},\,m_{-} $ are the masses on the boundaries of the balayage measure defined in \eqref{m-pm} below.

\item For the non-overlapping case $([\rho_-,\rho_+]\cap (r_0,r_1)= \emptyset$ and $\omega_-=\omega_+=\omega)$, the integral over $I$ vanishes and the sum of the two boundary terms in \eqref{ea-asy-thm} is instead, with $\tau_\omega :=\omega q'(\omega)/2$ (extension of \eqref{taur} beyond [0,1]),
\begin{align}
\mathcal{E}_-[a]+\mathcal{E}_+[a]=
\begin{cases}
\frac{\omega a'(\omega)}{2}\log\left|\frac{\tau_\omega}{\tau_\omega-1}\right|,
& \rho_->r_1\ \text{or}\ \rho_+<r_0,\\
\displaystyle
\frac{\omega a'(\omega)}{4}\left(\log N+\log\frac{2\pi}{\omega^2\Delta q(\omega)}\right),
& \rho_-=r_1,\\
\displaystyle
-\frac{\omega a'(\omega)}{4}\left(\log N+\log\frac{2\pi}{\omega^2\Delta q(\omega)}\right),
& \rho_+=r_0.
\end{cases}
\end{align}
\end{itemize}
}
\begin{remark}
    This theorem shows that the presence of hard walls leaves the quadratic term in $a(r)$ (corresponding to the bulk variance) $\frac14\int_{I}r\,a'(r)^2dr$ unchanged, while it produces endpoint dependent corrections for the centering which is determined by the one-point density. Given the prominence of fluctuation formulas in the study of random matrices, and more generally the one-component plasma
    (recall the references in the paragraph above \eqref{1.19}), these results motivate an investigation beyond the radially symmetric case, which is left for future research. Another noteworthy point is that when a hard wall lies strictly inside the unconstrained droplet, there is a structural change in the expansion \eqref{ea-asy-thm}, as the  correction then involves a term of order $\log N$.
\end{remark}

\begin{remark}
    The first edge term for the non-overlapping case ($[\rho_-,\rho_+]\cap (r_0,r_1)= \emptyset$) can be formally obtained using \eqref{E-pm} by letting both endpoints behave like strictly-inside hard walls; the latter two can be obtained by letting one of the endpoints behave like a  strictly-inside hard wall while the other like a critical hard wall.
\end{remark}
 
Up to and including the constant term, the large $N$ expansion of the term $\log Z_N^{\rm a}(q)$ in \eqref{b.1}
is known from \cite{BKS23} in the soft wall case, and from
\cite{AFLS25} in the presence of a single hard wall. By completing
the analysis of the missing two hard wall cases (albeit assumed to be entirely on the boundary, or inside of the droplet), we arrive at the following large $N$ expansion of $\log Z_N^{\rm c}(q)$ for any of the nine combinations of two soft/critical-hard/effective-hard walls. Substituting these results as appropriate in (\ref{b.1}) makes explicit the large $N$ form of $\log Z_N^{\rm c}(Q^{\rm c})$.

\theo{thm2hw}{
Let $\mu_q^{\Omega} $ be the equilibrium measure \eqref{muq-ome} with potential $q$ and confining interval $\Omega$. Let $I_{\mathbb{S}\cap\mathbb{A}}[\mu_q^{\Omega}],\,E_{\mathbb{S}\cap \mathbb{A}}^{\rm bulk}[\mu_{q}^{\Omega}],\,F_{\mathbb{S}\cap \mathbb{A}}[q]$, $\eta$ be as defined in \eqref{I-muq}, \eqref{E-muq}, \eqref{F1} 
and \eqref{eta} respectively.
 The large $N$ expansion of the logarithm of the partition function $\log {Z}_N^{\rm c}[q;\Omega]$ is given up to and including the ${\rm O}(1)$ term by
\begin{align}
    \begin{aligned}
	\log {{Z}_N^{\rm c}[q;\Omega]
    \over (2 \pi)^N}=-& N^{2}I_{\mathbb{S}\cap \mathbb{A}}[\mu_{q}^{\Omega}]-\frac{1}{2}(1+\eta)N\log N-\frac{N}{2}\left( E_{\mathbb{S}\cap \mathbb{A}}^{\rm bulk}[\mu_{q}^{\Omega}]+(1+\eta)\log 2+\mathcal{E}_{-}^{(N)}+\mathcal{E}_{+}^{(N)} \right) \\
	&-\left( \mathcal{E}_{-}^{(\sqrt{ N })}+ \mathcal{E}_{+}^{(\sqrt{ N })} \right) \sqrt{ N }+F_{\mathbb{S}\cap \mathbb{A}}[q]+\mathcal{E}_{-}^{(1)}+\mathcal{E}_{+}^{(1)}+{\mathrm{o}(1)}.
	\end{aligned}
\end{align}
Here $\mathcal{E}_{\pm}^{(N)},\mathcal{E}_{\pm}^{(\sqrt{ N })},\mathcal{E}_{\pm}^{(1)}$ are, for the specified 
upper ($+$) and lower ($-$) wall type, given by

{\centering
\footnotesize
\setlength{\tabcolsep}{2pt}
\renewcommand{\arraystretch}{1.5}
\begin{tabular}{c|c|c|c}
& soft wall & hard wall at edge & hard wall inside \\ \hline
$\mathcal{E}_{\pm}^{(N)}$
& $\mp V_{\tau_{\pm}}(\omega_{\pm})$
& $\mp V_{\tau_{\pm}}(\omega_{\pm})$
& $\begin{gathered}
\mp V_{\tau_{\pm}}(\omega_{\pm})
{}+2m_{\pm}(\log m_{\pm}-1)
\end{gathered}$\\ \hline
$\mathcal{E}_{\pm}^{(\sqrt{N})}$
& $0$
& $\omega_{\pm}\sqrt{\Delta q(\omega_{\pm})}\,\gamma^{\mathrm{in}}$
& $\omega_{\pm}\sqrt{\Delta q(\omega_{\pm})}
(\gamma^{\mathrm{in}}+\gamma^{\mathrm{out}})$\\ \hline
$\mathcal{E}_{\pm}^{(1)}$
& $\displaystyle
\pm\frac14\log\!\big(\omega_{\pm}\Delta q(\omega_{\pm})\big)$
& $\begin{gathered}
\displaystyle\pm\frac14\log\!\big(\omega_{\pm}\Delta q(\omega_{\pm})\big)\mp\tilde{\alpha}^{\mathrm{in}}\\
\displaystyle{}\pm
\frac{\omega_{\pm}\partial_r\Delta q(\omega_{\pm})}
{\Delta q(\omega_{\pm})}\,\beta^{\mathrm{in}}\pm \frac{1}{2}\log 2
\end{gathered}$
& $\begin{gathered}
\displaystyle\pm\frac14\log\!\big(\omega_{\pm}\Delta q(\omega_{\pm})\big)
\displaystyle{}\pm
\frac{\omega_{\pm}\partial_r\Delta q(\omega_{\pm})}
{\Delta q(\omega_{\pm})}
(\beta^{\mathrm{in}}+\beta^{\mathrm{out}})\\
\displaystyle{}\pm
\left(\tilde{\alpha}^{\mathrm{out}}-\tilde{\alpha}^{\mathrm{in}}+\frac14\log\pi\right)
+\frac{\omega_{\pm}^{2}\Delta q(\omega_{\pm})}{4m_{\pm}}\pm \frac{1}{2}\log 2
\end{gathered}$\\ \hline
\end{tabular}\par}

In the table the constants $\tilde{\alpha}^{\rm in},\beta^{\rm in},\gamma^{\rm in},\tilde{\alpha}^{\rm out},\beta^{\rm out},\gamma^{\rm out}$ are defined in \eqref{constin} and \eqref{constout}. The droplet boundaries $\omega_{\pm}$ and the masses $m_{\pm},\tau_{\pm}$ are given in \eqref{ome+-} and \eqref{m-pm}. $V_{\tau}(r)$ is defined in \eqref{3.3}.
}
\begin{proof}
    See Appendix~\ref{appendixB}.
\end{proof}

\begin{remark}\label{R1.2a}
It has been commented that the general $\beta$ free energy expansion for the cylinder (\ref{0.11}) can be anticipated to not have a term proportional to $\log N$
 due to the Euler characteristic of a cylinder being zero --- recall the text below \eqref{0.11a}.
 Our exact results for $\beta = 2$ suggest that this remains true in the presence of hard walls at the boundary of, or inside of, the droplet. If so, this universality is more robust than the appearance of the additive term ${\pi L \over 6 W}$ which requires a constant background density, and no hard wall inside of the droplet; see also Remark \ref{R1.6}.
 \end{remark}
 
\begin{remark}\label{R1.2}
The asymptotic expression with an effective lower hard wall is seen to be independent of $r_0$. It can
be checked to remain valid in the case $r_0=0$, when the
droplet is a disk rather than an annulus.
\end{remark}

\begin{remark}
    The ${\rm O}(N)$ edge terms $\mathcal{E}_{\pm}^{(N)}$ can be interpreted as a summation of the logarithmic potential and edge entropy. Taking the arclength measure as the reference measure, one can define the edge entropy as 
\begin{align*}
    	\begin{aligned}
		E_{\mathbb{S}\cap\mathbb{A}}^{\rm edge}[\mu_q^{\Omega}]:=&\int_{0}^{2\pi} \log \left( \frac{m_{-}}{2\pi \omega_{-}} \right)  \cdot  \frac{m_{-}d\theta}{2\pi}+\int_{0}^{2\pi} \log \left( \frac{m_{+}}{2\pi \omega_{+}} \right)  \cdot  \frac{m_{+}d\theta}{2\pi}\\
		=& m_{+}\log \frac{m_{+}}{\omega_{+}}+m_{-}\log \frac{m_{-}}{\omega_{-}}-\eta \log (2\pi).
	\end{aligned}
\end{align*}
Consider the logarithmic potential
\begin{align*}
    	\begin{aligned}
U_{\mu_q^{\Omega}}(z)
=&\int  \log  \frac{1}{|z-\omega|}d\mu_{q}^{\Omega}(\omega){}\\
=&-\frac{1}{2}\int_{\omega_{-}}^{\omega_{+}}s\Delta q(s)
\log\max(r,s)\,ds-\tau_-\log\max(r,\omega_{-})-(1-\tau_{+})\log\max(r,\omega_{+}).
\end{aligned}
\end{align*}
Direct calculation shows
\begin{align*}
    	U_{\mu_q^{\Omega}}(0)= \frac{1}{2}\left( V_{\tau_{+}}(\omega_{+})-V_{\tau_{-}}(\omega_{-}) \right)-m_{+}\log \omega_{+}-m_{-}\log \omega_{-}.
\end{align*}
Therefore, the edge terms can be expressed as
$$
\mathcal{E}_{-}^{(N)}+\mathcal{E}_{+}^{(N)}=2(E_{\mathbb{S}\cap\mathbb{A}}^{\rm edge}[\mu_q^{\Omega}]-U_{\mu_q^{\Omega}}(0))+2\eta(\log(2\pi)-1).
$$
\end{remark}

\begin{remark}
    By combining Theorems \ref{thm1} and \ref{thm2hw}, one obtains the asymptotics of the standard random normal matrix partition function with two hard walls for a general subharmonic rotationally invariant potential $q$, according to the second equality in \eqref{b.1} and \eqref{eq9}. Recalling Remark \ref{R1.2},
    this generalizes the $g=2$ case of \cite[Theorem~1.9]{Ch24}. In particular, for the hard walls strictly inside the droplet, there is no $\log N$ term, and similarly if both hard walls are on the droplet boundary. On the other hand, if one hard wall is inside the droplet, and the other at the boundary, there is a $\log N$ term.
    We make a consistency check with the latter  in \S \ref{S5.2}.
\end{remark}

 As an illustration of Theorem \ref{thm2hw}, we will note here some explicit forms upon specialising to the potential
$Q^{\rm c}(Y) = {W \over L} Y^2$, which corresponds to the case of a constant background $\rho_{\rm b} = N/(WL)$ in the setup relating to
(\ref{0.11}); see \S \ref{S2.2.1} for more details.
According to \eqref{eq5}  we then have
\begin{equation}\label{eq6}
q(r) = {W \over 2 \pi L} (\log r)^2 + \log r,  \quad
q'(r) = {W \over  \pi L} {\log r \over r} + {1 \over r},
\quad \Delta q(r) = {1 \over r} (r q'(r))' =
{W \over  \pi L r^2}.
\end{equation}
Making use of (\ref{rr01}) to determine the inner $r_0$ and outer $r_1$ radii of the droplet support shows
\begin{equation}\label{eq7}
 r_0 = e^{- \pi L/ W}, \qquad r_1 = e^{ \pi L/ W}.
\end{equation}
(It can be checked that
these formulas for $r_0$ and $r_1$ are consistent with the first formula in \eqref{eq4A} below, given that $ W Y_l \in [- {L \over 2}, {L \over 2}]$.) 
To focus the attention on the particular constant term in \eqref{0.11}, 
we restrict consideration 
to soft edge boundary conditions, one hard wall at the droplet boundary and one soft wall, or two hard walls both positioned at the droplet boundary. 

\propo{P1.1}{Let $Q^{\rm c}(Y) = {W \over L} Y^2$.
 For the case of soft walls,
\begin{equation}\label{b.4}
\log Z_N^{\rm c}(Q^{\rm c}) 
= {\pi L \over 6 W}N^2 - {1 \over 2} 
N \log N + N \bigg ( \log  (  2^{1/2} \pi   ) + {1 \over 2} \log {L \over W} \bigg ) - {\pi L \over 6W}
\end{equation}
exactly. 
Denote the right hand side of (\ref{b.4}) by $ \log Z_N^{\rm sc}(Q^{\rm c})$.
For the case of a hard wall at each of the droplet boundaries $Y = \pm {L \over 2 W}$,
\begin{equation}\label{b.5}
\log Z_N^{\rm c}(Q^{\rm c}) = 
\log Z_N^{\rm sc}(Q^{\rm  c}) 
+ 2 W  \gamma_{\partial \Omega}
 + {\rm o}(1),
\end{equation}
where 
$\gamma_{\partial \Omega}$ is given by (\ref{0.11b}), and as such is proportional to $\sqrt{N}$. (Note that according to (\ref{1.10a}), $W  \gamma_{\partial \Omega} = \sqrt{N W \over \pi L} \gamma^{\rm in}$, where $\gamma^{\rm in}$ is a numerical constant.)
For a hard wall placed at only one of these boundaries, with the charges otherwise free to occupy the droplet support,
\begin{equation}\label{b.6}
\log Z_N^{\rm c}(Q^{\rm c}) = 
\log Z_N^{\rm sc}(Q^{\rm  c}) 
+  W  \gamma_{\partial \Omega}
 + {\rm o}(1).
\end{equation}
}

\begin{proof} See \S \ref{S5.1}.
\end{proof}

\begin{remark}\label{R1.6}
The results of Proposition \ref{P1.1} show that the effect of placing a hard wall at a boundary of the droplet is restricted to creating an otherwise absent  surface tension term
(a peculiarity of soft walls at $\beta = 2$; recall (\ref{0.11a})). In particular, the universal constant term in \eqref{0.11} is unaltered. On the other hand, with the hard wall(s) placed inside the droplet, this is no longer true, as can be seen from the formulas of Theorem \ref{thm2hw}. In fact, a complicated function of $L/W$ then results, rather than one that is linear in this ratio as exhibited in
Proposition \ref{P1.1}.
\end{remark}

In Appendix \ref{AppD} we further consider the 
large $N$ form of the partition function in the case of the potential $Q^{\rm c}(Y)=WY^2/L$, now with the
plasma confined to the outside of two hard walls positioned in the interior of the droplet. This relates to the so-called gap probability, whereas the expansion
of $\log Z_N^{\rm c}(Q^{\rm c})$ with two hard walls in Theorem \ref{thm2hw}
relates to what is termed an overcrowding probability; see e.g.~\cite{AS13}.
A feature in common with the above asymptotic expansions is the absence of a $\log N$ term. A feature in distinction is that the constant term is an oscillatory function in $N$. We remark oscillatory terms have also been found in the study \cite{Ch24} relating to gap probabilities for (generalisations of) the one component plasma with a constant background with radial symmetry.

\subsection{Organisation of the paper}
In \S\ref{S2.2.1} 
we make explicit the change of variables \eqref{eq4A} that maps the cylinder OCP at $\beta=2$ to a planar normal matrix model with a radially symmetric potential $q(r)$ \eqref{eq5}, supported on an annulus, times an extra Jacobian factor $\prod_l |z_l|^{-1}$. In \S\ref{S2.2.2}
we give preliminaries relating to the Laplace asymptotic expansion of the definite integrals determining \eqref{1.19}. The analysis for this is completed in Section~\ref{sec:ratio}.
A second necessary asymptotic analysis, as required by the formulas for the logarithm of the partition function obtained in  \S\ref{S2.2.1}, is that of expanding a sum for large $N$, 
 for which use is made of the Euler--Maclaurin summation formula. As such we are following the framework developed in \cite{BKS23}, \cite{AFLS25} for the radial planar case. 
In \S \ref{S5} we 
give the proof of Proposition \ref{P1.1}, and further
specialise our results for the asymptotic expansion of the cylinder partition in annular coordinates to the particular potential $q(r) = r^{2\mathfrak b}$. This latter special case has been treated earlier in \cite{Ch24}, providing an opportunity to
check our results.
Three appendices collect technical material: 
Appendix \ref{appendix0} is on the derivation of the pair potential for the Coulomb potential on the surface of a cylinder; Appendix~\ref{appendixA} records the Euler--Maclaurin formula used throughout, and Appendix~\ref{appendixB} gives the proof of Theorem \ref{thm2hw} in the case of two hard walls inside the droplet. A fourth Appendix 
provides an asymptotic expansion relating to the gap probability in the cylinder model with a constant background.

\section{Coulomb gas on cylinder --- formalism facilitating  large $N$ analysis}\label{sec:cylinder}

\subsection{Mapping from the  cylinder droplet to the annular droplet Boltzmann factor}\label{S2.2.1}

For the cylinder viewed as a semi-periodic strip, period $W$ in the $x$-direction, it is shown in Appendix A that the pair potential is given by (\ref{0.6}).
For a system of $N$ particles, with a neutralising background chosen to be rectangle $x \in [0,W]$, $y \in [-L/2,L/2]$, the corresponding Boltzmann factor is known to be \cite{CFS83},
\cite[Eq.~(2.8)]{CFTW15}
\begin{equation}\label{eq1}
A_{N,\beta}^{\rm c} e^{- \pi \beta \rho_{\rm b} \sum_{j=1}^N y_j^2}
\prod_{1 \le j < k \le N} \Big | 2 \sin { \pi (x_j - x_k + i (y_j - y_k)) \over W} \Big |^\beta, \quad   A_{N,\beta}^{\rm c}  = \Big ( {W \over 2 \pi} \Big )^{-N \beta/2} e^{- \pi \beta N L^2 \rho_{\rm b} / 12},
\end{equation}
where $\rho_{\rm b}=N/LW$ is the background density.
For $\beta$ even and $w_j = (x_j + i y_j)/W$ we have \cite[displayed equation below (3.16) with a correction]{CFTW15}
$$
 \Big | 2 \sin { \pi (x_j - x_k + i (y_j - y_k)) \over W} \Big |^\beta = e^{\pi \beta (y_j + y_k)/W} ( e^{2 \pi i w_j} - e^{2 \pi i w_k} )^{\beta/2} ( e^{-2 \pi i \bar{w}_j} - e^{-2 \pi i \bar{w}_k} )^{\beta/2},
$$
which implies the alternative form of the Boltzmann factor
\begin{equation}\label{eq1B}
A_{N,\beta}^{\rm c} \prod_{j=1}^{N} e^{\pi \beta ((N-1)y_j/W-\rho_by_j^2)}\prod_{1 \le j < k \le N} \Big |  e^{2 \pi i w_j} - e^{2 \pi i w_k}  \Big |^{\beta}.
\end{equation}
Note that repeating this working with $\pi \rho_{\rm b} \sum_{j=1}^N y_j^2$ replaced by $N \sum_{j=1}Q(-y_j)/2$, 
$y_j \mapsto - y_j$ and then setting $\beta = 2$
and changing variables $X_j = x_j/W$, $Y_j = y_j/W$, we obtain (\ref{eq4}) up to the prefactor. In this latter expression, the change of variables
\begin{equation}\label{eq4A}
e^{2 \pi Y_l} =  r_l, \qquad  2 \pi e^{2 \pi Y_l} dY_l =  d r_l.
\end{equation}
gives
\begin{equation}\label{eq4B}
 \prod_{l=1}^N e^{-2 \pi N  Q^{\rm c} ((1/2 \pi) \log  r_l  )} r_l^{-N} 
 \prod_{1 \le j < k \le N} \Big |  r_j e^{2 \pi i X_j} - r_k e^{2 \pi i X_k}  \Big |^2.
\end{equation}
 Regarding each $X_l$ as a scaled angle, we see that
 $r_l e^{2 \pi i X_l} =: r_l e^{i \theta_l}$ corresponds to the polar form of a complex number $z_l$. Consideration of the corresponding Jacobian gives the transformation of
 (\ref{eq4B})
\begin{equation}\label{eq5A}
 \Big ( {1 \over 2 \pi } \Big )^{N}
 \prod_{l=1}^N e^{-N  q(|z_l|)} |z_l|^{-1}  
 \prod_{1 \le j < k \le N}  |  z_j - z_k    |^2,
\end{equation}
where $q(r)$ is specified by (\ref{eq5}).
Denote the partition function for \eqref{eq5A} without the factor of $\prod_{l=1}^N |z_l|^{-1}$ by $(2 \pi)^{-N}{Z}_N^{\rm a}(q)$
(this is in keeping with (\ref{1.9}) and the notation introduced at the beginning of the paragraph below (\ref{eq:ZW-conj})), 
and the partition function  with this factor by $(2 \pi)^{-N} {Z}_N^{\rm c}(q)$ which is consistent with the notation below (\ref{eq5}).
Considering the removed prefactor in the definition of 
$Z_N^{\rm c}(q)$, the equation (\ref{b.1}) follows. An essential point is that the large $N$ form of the 
logarithm of ${Z}_N^{\rm a}(q)$, up to and including the constant term, is known from \cite{BKS23} in the soft wall case and
\cite{AFLS25} in the case of a single hard wall. The product of differences in (\ref{eq5A}) with the absolute value to the power of $\beta$ removed is recognised as one side of the 
Vandermonde determinant identity (see e.g.~\cite[Eq.~(1.173)]{Fo10}). Specialise now to
$\beta = 2$ and substitute the determinant form for each of the two products of differences. One sees from the
Andr\'eief formula \cite{Fo19}, which in general reduces the integration over the product of two determinants,
$\det [f_j(z_k)]_{j,k=1}^N$ times
$\det [g_j(z_k)]_{j,k=1}^N$, to a single determinant, that after integrating over the angular variables, only the diagonal terms in the single determinant are nonzero.
Hence the multiple integral factorises as a product of one-dimensional integrals \cite{Ch81,CFS83}, and thus
\begin{equation}\label{logratioZ}
\log {{Z}_N^{\rm a}(q) \over (2 \pi)^N}
= \sum_{l=0}^{N-1} \log u_l^{\rm a}, \quad  \log {{Z}_N^{\rm c}(q) \over (2 \pi )^N} =\sum_{l=0}^{N-1}  \log u_l^{\rm c},
\end{equation}
where
\begin{equation}\label{eq10A}
u_l^{\rm a} = \int_0^\infty e^{-N q(r)} r^{2l+1} \, dr, \quad
u_l^{\rm c} = \int_0^\infty e^{-N q(r)} r^{2l} \, dr.
\end{equation}

\subsection{The extended integral $u_j[a]$ and basic settings}\label{S2.2.2}
To extend our analysis beyond (\ref{logratioZ}), and to shed light on the hard-wall effect for the Gaussian fluctuation formula, we focus on the form 
(\ref{eq9}) and consider the more general average
$	\left< e^{ \sum_{i}a(\lvert z_{i} \rvert ) } \right>_{\mathrm{a}}$, 
where $a(r)$ is assumed to be analytic on and in a neighbourhood of the droplet. This in turn requires considering the
generalisation of the integrals in (\ref{eq10A})
\begin{align}\label{uja}
    	u_{j}[a]:=\int_{\Omega} r^{2j+1} e^{a(r)}\,e^{-Nq(r)}\,dr,
\end{align}
with $a(r)=0,-\log r$ corresponding to $u_j^{\rm a},\,u_{j}^{\rm c} $ respectively. 
As in \cite{BKS23}, \cite{AFLS25} it is to be assumed that
$q$ is $C^{\infty}$, subharmonic in $\mathbb C$, and strictly subharmonic in a neighbourhood of the droplet. The latter conditions imply that $\Delta q(r)=\frac{1}{r}(rq'(r))'>0$ in a neighbourhood of the droplet, which tells us that away from the origin, $rq'(r)$ is strictly increasing in the same neighbourhood.
We are interested in the large $N$ expansion for three cases distinguished by the number of hard walls (0,1,2). The integration region in the definition of $u_{j}[a]$ should also be adjusted accordingly by the restriction of the hard wall(s). Denote 
\begin{align}
    \Omega=(\rho_{-},\rho_{+}),
\qquad
0\leq \rho_{-}<\rho_{+}\leq\infty,
\end{align}
then all wall configurations are included in the choices
\begin{align*}
    \begin{array}{c|c}
\Omega & \text{physical confinement} \\
\hline
(0,\infty) & \text{no hard wall},\\
(\rho_{-},\infty) & \text{one inner hard wall},\\
(0,\rho_{+}) & \text{one outer hard wall},\\
(\rho_{-},\rho_{+}) & \text{two hard walls.}
\end{array}
\end{align*}
It is well known that under mild assumptions, the empirical measure $\frac{1}{N}\sum_{i=1}^{N}\delta_{z_{i}}$ of the model \eqref{eq5A} --- excluding the factor $\prod_{l=1}^N |z_l|^{-1}$ which in fact is irrelevant from the viewpoint of this observable --- weakly converges to an equilibrium measure $d\mu_q^{\Omega}(z)$. For the soft wall case, the measure takes the form 
\begin{align}
    d\mu_q^{(0,\infty)}(z)=\frac{\Delta q(|z|)}{4}\chi_{\mathbb{S}}\,dA(z),
\end{align}
where $\mathbb{S}=\left\{z \in \mathbb{C}: r_0 \leq|z| \leq r_1\right\}$ is the corresponding droplet. Here $r_0$ is the largest solution to $r q^{\prime}(r)=0$ and $r_1$ is the smallest solution to $r q^{\prime}(r)=2$.
In the presence of effective hard walls, the droplet is restricted by the hard walls. Denote by 
\begin{align}\label{mathbbA}
    \mathbb{A}=\left\{z \in \mathbb{C}:\,|z|\in \bar{\Omega} \right\} 
\end{align}
the hard wall restriction and let $I=[\omega_-,\omega_+]$ with $\omega_{\pm}$ defined in \eqref{ome+-}.
Then for all hard wall cases, $d\mu_q^{\Omega}(z)$ can be expressed in a unified way
\begin{align}\label{muq-ome}
    d\mu_q^{\Omega}(z)
=\frac{\Delta q(|z|)}{4}
\mathbf \chi_{\mathbb{S}\cap\mathbb{A}}\,dA(z)
+m_-\,d\sigma_{\omega_{-}}(z)
+m_+\,d\sigma_{\omega_{+}}(z),
\end{align}
where  $d\sigma_\omega(z)=\chi_{\{|z|=\omega\}} dz/(2\pi iz)$ denotes the normalized angular measure on the circle $|z|=\omega$. The boundary masses are given by
\begin{align}\label{m-pm}
    \begin{aligned}
		&m_{-}=\int_{|z|\in[r_{0},\omega_{-}]}d\mu_{q}^{(0,\infty)}(z)=\int_{r_{0}}^{\omega_{-}} \frac{r\Delta q(r)}{2}dr=\frac{\omega_{-}q'(\omega_{-})}{2}:=\tau_{-},\\
		& m_{+}=\int_{|z|\in[\omega_{+},r_{1}]}d\mu_{q}^{(0,\infty)}(z)=1-\frac{\omega_{+}q'(\omega_{+})}{2}:=1-\tau_{+.}
	\end{aligned}
\end{align}
It turns out that the radial mass variable
\begin{align}\label{taur}
    \tau_r=\tau(r):=\mu_q^\Omega\big(\{z:|z|\leq r\}\big)=\frac{rq'(r)}{2},\, r\in[r_0,r_1],
\end{align}
already evident in \eqref{m-pm},
plays a prominent role. Since the map is strictly increasing on the natural droplet $[r_0,r_1]$, we can take its inverse and denote it by $r_\tau$ (note that this is consistent with the definition of $r_0,r_1$). It is clear that the variable $\tau$ ranges over $[0,1]$, so that we may further associate it to the quantity $j/N,\,j=0,\dots,N-1$ within the same range. From now on, we adopt the convention that $\tau$ or $\tau(j)$ refers to the independent variable obtained from $j/N$ by allowing $j$ to take continuous values, while $\tau_r$ refers to the quantity defined in \eqref{taur}.
Under this notation, we rewrite the integrals in \eqref{uja} as 
\begin{align*}
    u_{j}[a]=\int_{\Omega} G(r)e^{-N V_{\tau(j)}(r)}dr,
\end{align*}
where $G(r)=re^{a(r)} $ and $V_\tau(r)$ take the form
\begin{equation}\label{3.3}
    V_{\tau}(r):=q(r)-2\tau \log(r).
\end{equation}
Analysing the large $N$ asymptotics of the integral $u_{j}[a]$ using Laplace's method requires knowledge of the location of the minima of $V_\tau(r)$. The critical point equation 
\begin{align}
    V_{\tau}'(r)=q'(r)-\frac{2\tau}{r}=0
\end{align}
shows that for each $\tau$, the inverse map $r_\tau$ of \eqref{taur} gives the unique minimum point inside the droplet, i.e., it satisfies the relations
\begin{align*}
    V_{\tau}'(r_\tau)=0,\quad (r-r_\tau)V'_\tau(r_\tau)\geq 0,\quad V_\tau''(r_\tau)>0,\quad r\in[r_0,r_1].
\end{align*}
The presence of the hard walls is then encoded by $\tau(\rho_\pm)\in[0,1]$, and the relative position of the minimum point and the hard walls is determined through
\begin{align}
    \begin{cases}
        \tau<\tau(\rho_-) \text{ or } \tau>\tau(\rho_+),\quad &\text{$r_\tau$ is outside the hard walls},\\
        \tau(\rho_-)<\tau<\tau(\rho_+),\quad & r_\tau \text{ is inside the hard walls}.
    \end{cases}
\end{align}
The expression of $r_\tau$ is implicit and so is its derivative with respect to $\tau$. However, one can relate the two by using implicit differentiation on \eqref{taur} to arrive at
\Eq{eq: r'}{
    \frac{d r_\tau}{d\tau}=\frac{2}{ r_\tau \Delta q(r_\tau)}.
}
For later use, using the equation $\Delta q(r)=\frac{1}{r}(rq'(r))' $ and \eqref{taur}, the derivatives of $V_{\tau}(r) $ can be expressed as
\begin{align}\label{deriV}
    \begin{aligned}
        &V_\tau''(r)=\Delta q(r)-\frac{1}{r}V_\tau'(r),\quad  V_\tau'''(r)=\partial_r \Delta q(r)-\frac{1}{r}\Delta q(r)+\frac{2}{r^2}V'_\tau(r),\\
    &V_\tau^{(4)}(r)=\partial_r^2 \Delta q(r)-\frac{1}{r}\partial_r\Delta q(r)+\frac{3}{r^2}\Delta q(r)-\frac{6}{r^4}V'_\tau(r), \\
    &V'_\tau(r)=\frac{2}{r}(\tau_r-\tau).
    \end{aligned}
\end{align}

Let 
$$
	Z_{N}[q,a;\Omega]:=\sum_{j=0}^{N-1}\log u_{j}[a]
$$
be the partition function associated to \eqref{uja}. It is possible to rescale the functions to fix the outer hard wall at $r=1$ and consider the alternative
\begin{align}\label{ZN-sca}
    	Z_{N}[\widehat{q},\widehat{a};(b,1)]=Z_{N}[q,a;(\rho_{-},\rho_{+})]-N^{2}\log \rho_{+},
\end{align}
where $\widehat{q}(r)=q(\rho_+r),\,\widehat{a}(r)=a(\rho_+r) $ and $b=\rho_-/\rho_+$. Note that $\Delta\widehat{q}(r)=\rho_+^2\Delta q(\rho_+r)$ so the assumptions on $q$ are unchanged under the scaling. In 
Sections 3, 4 and Appendix C, subsections C.1 on, we will assume the hard walls are scaled to $r=b,1$ and abbreviate the notation to $Z_N^{\rm c}(q),\,Z_N^{\rm a}(q) $ for the cylinder and annulus cases, respectively.

\section{The ratio of partition functions and Laplace's method}\label{sec:ratio}


Following \cite{BKS23}, \cite{AFLS25},
the strategy to be used for the asymptotic expansion of \eqref{uja} is to first apply Laplace's method to gain sufficient control of the asymptotics of the integral $u_{j}[a]$. 
For use in relation to the Gaussian fluctuation formula,
this will then be used to determine the asymptotics of the required sums (analogous to those in \eqref{logratioZ}) using the Euler-Maclaurin formula. The first of these tasks will be addressed in the 
remainder of this section, while the second is the topic
of \S \ref{S4}.

\subsection{Asymptotic expansion of the summands $u_{j}[a]$ --- introductory remarks}

As shown in the earlier work \cite{AFLS25}, the asymptotic analysis is different depending on whether $r_{\tau}$ is inside or outside the hard walls, which will be treated separately. Moreover, we need to introduce two scalings on $\tau$ to separate the cases that $r_{\tau} $ is close to the hard walls or away from the hard walls. When $r_{\tau}$ approaches one of the hard walls from outside, for some $\epsilon>0$ we impose a scaling $\Delta_N $ that satisfies
\begin{align}\label{Delta_N}
    \Delta_N\sim {\rm O}(N^{-1/4+\epsilon}),\quad \text{and } \Delta_N \to 0, \quad \sqrt{N} \Delta_N \to \infty \quad \text{as } N \to \infty.
\end{align}
When $r_{\tau} $ approaches the hard walls from inside, the corresponding scaling $\delta_N$ is set to be
\begin{align}\label{delta_N}
    \delta_N := \frac{(\log N)^2}{\sqrt{N}}.
\end{align}

To treat the two hard walls in a unified way, we introduce some further notations. For a wall $\omega\in\{b,1\}$, let $\tau_\omega\in\{\tau_b,\tau_0\}$ and let $\varsigma_\omega$ be its outward orientation in $\tau$:
\begin{align*}
    \varsigma_b:=-1,\qquad \varsigma_1:=+1,
\end{align*}
so that $t_\omega:=\varsigma_\omega (\tau-\tau_\omega )$ is positive when $r_\tau$ lies outside the wall and negative inside. For the cases when the stationary point $r_{\tau}$ is close to the hard walls characterised by the scalings defined above ($0<t_{\omega}<\Delta_N $ for the outside and $0<-t_\omega<\delta_N $ for the inside case), we will need a boundary-layer variable at either wall
\begin{align}\label{x-ome}
    x_\omega :=\frac{1}{\omega }\sqrt{\frac{2N}{\Delta q(\omega )}}\;t_\omega ,\quad \omega\in\{b,1\}.
\end{align}

\subsection{The case $r_\tau$ inside the hard walls ($\tau\in (\tau_b,\tau_0)$).}
In this subsection, we derive the asymptotic expansions for $u_{j}[a]$ in the case that the subscript $j$ is within the range such that the critical point is inside the hard walls, i.e. $r_{\tau}\in [b,1]$. We start with the following proposition.
\propo{prop: uj asym4}{For $0<-t_\omega <\delta_N$, the integral $u_{j}[a]$ \eqref{uja} admits the $N\to\infty$ expansion
\begin{align}\label{eq: uj asym4}
     	u_j[a] =  \frac{\sqrt{2\pi}\,r_{\tau}\,e^{ a(r_{\tau}) }}{\sqrt{N V_\tau''(r_\tau)}}\,e^{-N V_\tau(r_\tau)}\left(f_2(x_{\omega})+ \frac{1}{\sqrt{N}}\frac{\varsigma_\omega }{\omega}\sqrt{ \frac{2 }{\Delta q(\omega)} }\,h_2^{a}(x_{\omega};\omega)+\mathrm{O}(N^{-1})\right).
\end{align}
Here $x_\omega$ is defined in \eqref{x-ome} and the functions $f_2$ and $h_2^a$ are given by
\begin{align}\label{eq:f2h2}
    \begin{aligned}
        &f_2(x):=\frac{1}{2}\erfc(x), \quad
h_2(x;\omega):=\frac{e^{-x^2} \left(2x^2-1\right)}{6 \sqrt{\pi }}+\frac{e^{-x^2} \left(x^2+1\right) }{6 \sqrt{\pi } }\frac{\omega\partial_r \Delta q(\omega)}{\Delta q(\omega)},\\
& h_2^{a}(x;\omega):=h_2(x;\omega)-\frac{e^{-x^{2}}}{2\sqrt\pi}\,\omega\,a'(\omega)\;.
    \end{aligned}
\end{align}
}
\begin{proof}
    As a special case, the asymptotic expansion for $u_j[0]$ has been treated in \cite{AFLS25}. The one for general $a$ can be proved in a similar manner. For completeness, we state the proof here. We treat only the case near the hard wall $r=b$. The condition $0<\tau-\tau_b<\delta_N$ implies that the stationary point $r_\tau$ is inside the hard walls and close to the lower boundary $r=b$. With the cut off scaling
    \begin{align}\label{beta_n}
        \beta_N=\frac{\log N}{\sqrt{N}},
    \end{align}
    we split the integral
    \begin{align*}
        u_{j}[a]=I_1+I_2,\quad I_1:=\int_{b}^{r_\tau+\beta_N}G(r)e^{-N V_{\tau}(r)}\, dr, \quad 
    I_{2} := \int_{r_\tau + \beta_N}^{1} G(r)e^{-N V_{\tau}(r)}\, dr.
    \end{align*}
    The integral $I_1$, which is on the interval including the local minimum point $r_\tau$ of the function $V_\tau(r)$, gives the main contribution to the $N\to \infty$ asymptotic expansion of $u_{j}[a]$. $I_2$ can be shown to be exponentially suppressed compared to $I_1$ in the following estimate. Recall that $V_\tau(r)$ is strictly increasing when $r>r_\tau$. It follows by a Taylor expansion at $r_\tau$ that 
    \begin{align*}
        N\left(V_\tau(r) - V_\tau(r_\tau)\right)
    \geq N\left(V_\tau(r_\tau + \beta_N) - V_\tau(r_\tau)\right)
    \geq C (\log(N))^2,
    \end{align*}
    for some constant $C > 0$.  This implies the bound
\begin{equation}
     I_{2}
    \leq e^{-N V_\tau(r_\tau)} \, e^{-C (\log(N))^2}.
    \end{equation}
    Let us turn to $I_1$. Expanding the exponent $V_\tau(r)$ in a Taylor series at $r_\tau$ and substituting it in the integral, with the change of variable $r=r_\tau+\frac{t}{\sqrt{N}}$, gives
    {\small\begin{align*}
        I_1\underset{N\to\infty}{=}\frac{G (r_\tau)e^{-N V_\tau(r_\tau)}}{\sqrt{N}}\int_{(b - r_\tau) \sqrt{N}}^{\log(N)}
     e^{-\frac{V_\tau''(r_\tau) t^2}{2}}\left(
        1 +\frac{(G )'}{G }(r_\tau)\,\frac{t}{\sqrt N} - \frac{1}{6} \sqrt{\frac{1}{N}} V_\tau^{(3)}(r_\tau) t^3
        + {\rm O}(N^{-1})
    \right)
     \, dt.
    \end{align*}}
    Since $V_\tau''(r_\tau)>0$ we can extend the integration domain from $[(b - r_\tau) \sqrt{N},\log(N)]$ to $[(b - r_\tau) \sqrt{N},\infty)$ with an exponentially small error. The integral can be computed explicitly and gives 
    \begin{align*}
    \sqrt{\frac{2\pi}{NV_\tau''(r_\tau)}}\,G (r_\tau)\,e^{-NV_\tau(r_\tau)}
	\left[\frac12\operatorname{erfc}(y_b)
	+\frac{e^{-y_b^{2}}}{\sqrt{2\pi N V_\tau''(r_\tau)}}
	\left(\frac{(G )'}{G }(r_\tau)-\frac{V_\tau'''(r_\tau)}{3V_\tau''(r_\tau)}(y_b^{2}+1)\right)\right].
    \end{align*}
    where
    \begin{align*}
        y_b=\frac{1}{\sqrt{2}}\sqrt{NV_\tau''(r_\tau)}(b-r_\tau).
    \end{align*}
    The claim \eqref{eq: uj asym4} for $u_j[a]$ follows from a further expansion with respect to 
    \begin{align*}
        r_\tau=b+\frac{2}{b\Delta q(b)}(\tau-\tau_b)-\frac{2(\Delta q(b)+b\partial\Delta q(b))}{b^3\Delta q(b)^3} (\tau-\tau_b)^2+{\rm O} (\tau-\tau_b)^3.
    \end{align*}
\end{proof}

\coro{}{For $0<-t_w<\delta_N$, the logarithm of the ratio ${u_j[a]}/{u_j[0]}$ admits the following asymptotic expansion
\begin{equation}
    	\log\frac{u_j [a]}{u_j [0]}
	= a(r_\tau)-\varsigma_\omega\frac{a'(\omega)}{\sqrt N}\sqrt{\frac{2}{\Delta q(\omega)}}\;\frac{e^{-x_\omega^{2}}}{2\sqrt\pi\, f_2(x_\omega)}+\mathrm{O}(N^{-1}),
\end{equation}
where $x_\omega$ and $f_2$ are given in \eqref{x-ome} and \eqref{eq:f2h2} respectively.
}

In the case that the local minimum point $r_\tau$ is inside the hard walls and away from the boundaries $r=1,b$, the asymptotics for $u_{j}[a]$ are given by the following proposition.
\propo{}{
If $\tau_b+\delta_N<\tau<\tau_0-\delta_N$, the integral $u_{j}[a]$ admits the $N\to \infty$ expansion
\Eq{asymp-uj-3}{
	u_j[a]\underset{N\to\infty}{=}\frac{\sqrt{2\pi}\,e^{a(r_\tau)}}{\sqrt{N V_\tau''(r_\tau)}}\,e^{-NV_\tau(r_\tau)}\left(1+ \mathcal{B}^{a}(r_\tau)\frac{1}{N}+\mathrm{O}(N^{-2})\right),
}
where
\Eq{Bc}{
&\mathcal{B}^a(r)=\mathcal{B}(r)+\frac{a''(r)+a'(r)^{2}}{2\Delta q(r)}
	+\frac{3a'(r)}{2r\,\Delta q(r)}
	-\frac{a'(r)\,\partial_r\Delta q(r)}{2\Delta q(r)^{2}},\\
&\mathcal{B}(r)=\frac{1}{3}\frac{1}{r^2\Delta q(r)}-\frac{19}{24 r} \frac{\partial_r \Delta q(r)}{\left(\Delta q(r)\right)^2} 
-\frac{1}{8} \frac{\partial_r^2 \Delta q(r)}{\left(\Delta q(r)\right)^2} 
+\frac{5}{24}\frac{\left(\partial_r \Delta q(r)\right)^2}{\left(\Delta q(r)\right)^3}.
}
}

\begin{proof}
   We keep the same truncation scaling $\beta_N$ as defined in \eqref{beta_n}. In the Taylor expansion 
   \begin{align*}
       r_\tau-r_{\tau'}=\frac{2}{r_{\tau'}\Delta q(r_{\tau'})}(\tau-\tau')+{\rm O}((\tau-\tau')^2), \quad \text{as } \tau\to \tau',
   \end{align*}
   by replacing $\tau'$ with $\tau_b,\tau_0$, we see that there exists some constant $c>0$ such that $r_\tau\in (b+c\delta_N,1-c\delta_N)$ for all $\tau\in(\tau_b+\delta_N,\tau_0-\delta_N)$. Since by definition, $\beta_N$ vanishes to a higher order than $\delta_N$ as $N\to \infty$, we have $(r_\tau-\beta_N,r_\tau+\beta_N)\subset (b,1)$ for sufficiently large $N$.
   Therefore, we separate the integral into three parts $u_j[a]=I_0+I_1+I_2$, where
   \begin{align*}
       I_0:=\int_{b}^{r_{\tau}-\beta_N}G(r)e^{-N V_{\tau}(r)}\, dr,\quad I_1:=\int_{r_{\tau}-\beta_N}^{r_\tau+\beta_N}G(r)e^{-N V_{\tau}(r)}\, dr, \quad 
    I_{2} := \int_{r_\tau + \beta_N}^{1}G(r) e^{-N V_{\tau}(r)}\, dr.
   \end{align*}
   The same estimate as carried out in the proof of Proposition \ref{prop: uj asym4} shows that $I_2$ is exponentially suppressed. A similar estimate holds for $I_0$. Therefore, we focus on $I_1$. Substituting the Taylor expansion of $V_{\tau}(r) $ at $r_{\tau}$ up to the fourth order and making the change of variable $s=\sqrt{N}(r-r_{\tau}) $ yields
   \begin{align*}
       &\frac{G(r_{\tau}) e^{-NV_{\tau}(r_{\tau})}}{\sqrt{N}}\int_{-\log N}^{\log N} 
e^{-\frac{1}{2} V_\tau''(r_\tau) \, s^2}
(F(r_\tau,s)+{\rm O}(N^{-2}))
\, ds,\\
&F(r_\tau,s)=1+\frac{1}{\sqrt{N}}\left(\frac{s G'(r_\tau)}{G(r_\tau)}-\frac{V_\tau^{(3)}(r_\tau)}{6 } s^3\right) \\
&\qquad + 
\frac{1}{N}\left(\frac{(V_\tau^{(3)}(r_\tau))^2}{72} s^6 - \frac{V_\tau^{(4)}(r_\tau)}{24} s^4 -\frac{G'(r_\tau)V_\tau^{(3)}(r_\tau)}{6G(r_\tau)}s^4+\frac{G''(r_\tau)}{2G(r_\tau)}s^2\right).
   \end{align*}
   We can again extend the integration limits to $\pm\infty$ with an exponentially small error. By doing so, the integral can be explicitly calculated and gives
   {\small\begin{align*}
       \frac{\sqrt{2\pi}G(r_\tau)}{\sqrt{NV_\tau''(r_\tau)}}e^{-NV_\tau(r_\tau)} \bigg(1+\frac1N\left(\frac{G''(r_\tau)}{2G(r_\tau)V_\tau''(r_\tau)}-\frac{G'(r_\tau)V_\tau'''(r_\tau)}{2G(r_\tau) V_\tau''^{2}(r_\tau)}+\frac{5V_\tau'''^{2}(r_\tau)}{24V_\tau''^{3}(r_\tau)}-\frac{V_\tau^{(4)}(r_\tau)}{8V_\tau''^{2}(r_\tau)}\right)\bigg).
   \end{align*}}
   A further substitution of the derivatives \eqref{deriV} gives the desired results in \eqref{asymp-uj-3} for $u_j[a]$.
\end{proof}

As a result, we obtain as a corollary, 
\coro{}{If $\tau_b+\delta_N<\tau<\tau_0-\delta_N$, the log ratio $\log\frac{u_j[a]}{u_j[0]}$ admits the $N\to \infty$ expansion
\begin{equation}
  \log\frac{u_j[a]}{u_j[0]}
	= a(r_\tau)+\frac{1}{N}\left[\frac{a''+(a')^{2}}{2\Delta q}
	+\frac{3a'}{2r\Delta q}
	-\frac{a'\,\partial_r\Delta q}{2\Delta q^{2}}\right]_{r=r_\tau}+\mathrm{O}(N^{-2}).
\end{equation}
}
\subsection{The cases $r_\tau $ outside the hard walls ($\tau\in [0,\tau_b]\cup[\tau_0,1]$).}
\propo{}{
If \(  0<t_\omega <\Delta_N \), then the integral $u_j[a]$ \eqref{uja} admits the $N\to\infty$  expansion
\begin{equation}\label{eq: uj asymb2}
     u_j[a] = \frac{e^{-N V_\tau(\omega)}\omega e^{a(\omega)}}{N} \sqrt{ \frac{2 N}{\Delta q(\omega)} } \left( f_1(x) + \frac{\varsigma_\omega}{\sqrt{N}\omega}\sqrt{ \frac{2 }{\Delta q(\omega)} }  h_1^{a}(x_\omega;\omega) + {\rm O}(N^{-1}) \right),
\end{equation}
where $x_\omega$ is defined by \eqref{x-ome}, and the functions $f_1,h_1^a$ take the following form
\begin{align}\label{eq:f1h1}
    \begin{aligned}
        &f_1(x):=\frac{\sqrt{\pi}}{2}e^{x^2}\erfc(x),\, h_1^{a}(x;\omega):=h_1(x;\omega)+\omega\,a'(\omega)\Bigl(xf_1(x)-\tfrac12\Bigr), \\
    &h_1(x;\omega):=\frac{1}{3}\left[(x^2-2)+x(3-2x^2) f_1(x) \right]+\frac{1}{6}\frac{\omega\partial_r\Delta q(\omega)}{\Delta q(\omega)}\left[(x^2+1)-x(2x^2+3)f_1(x) \right].
    \end{aligned}
\end{align}
}

\begin{proof}
    We consider the case $\tau_b-\Delta_N<\tau<\tau_b$. Since $\tau<\tau_b$, the phase function takes its minimum at $r=b$, with a nonzero first derivative $V_{\tau}'(b) $. After Taylor expanding the exponent, it is unclear whether $NV_{\tau}'(b)(r-b) $ or $NV_{\tau}''(b)(r-b)^2 $ would be the leading term, which then requires introducing an interpolating scaling $c_N:=\sqrt{N}+|V_{\tau}'(b)|N $. The integral is then split into two parts
    \begin{align*}
        u_j[a]=I_1+I_2,\quad I_1=\int_{b}^{b+d_N}G(r)e^{-NV_{\tau}(r)}dr,\,I_2=\int_{b+d_N}^1G(r)e^{-NV_{\tau}(r)}dr,
    \end{align*}
    with the cut-off scaling
    \begin{align}
        d_N:=\frac{(\log N)^2}{c_N}.
    \end{align}
    The bound
    \begin{align*}
        N(V_{\tau}(r)-V_{\tau}(b))\geq N(V_{\tau}(b+d_N)-V_{\tau}(b))\geq C_1(\log N)^2
    \end{align*}
    holds for some positive constant $C_1>0$ and $r\in [b+d_N,1] $. Therefore, we have the estimate
    \begin{align*}
        I_2\leq e^{-NV_{\tau}(b)}e^{-C_1(\log N)^2},
    \end{align*}
    which is exponentially suppressed as $N\to \infty$. For $I_1$, substituting the Taylor expansion of $V_{\tau}(r)$ and $G(r)$ at $r=b$ and making a change of variable $s=c_N(r-b) $ gives
    \begin{align*}
        I_1&=\frac{e^{-NV_{\tau}(b)}}{c_N} \int_{0}^{(\log N)^2}\left(G(b)+G'(b)\frac{s}{c_N}+{\rm O}\bigg(\frac{(\log N)^2}{c_N^2}\bigg)\right)\\
        & \qquad \times e^{-\frac{N V_\tau'(b)}{c_N} s
    -\frac{N V_\tau''(b)}{2 c_N^2} s^2-\frac{N V_\tau'''(b)}{6 c_N^3} s^3+{\rm O}(N(\log N)^4/c_N^4)}
\, ds.
    \end{align*}
    The choice of $c_N$ ensures that $\frac{N V_\tau^{(3)}(b)}{6 c_N^3} s^3$ is of order ${\rm O}((\log N)^6/\sqrt{N})$ as $N\to \infty$, so a further expansion can be made to get
    {\small\begin{align*}
        I_1=\frac{G(b)e^{-NV_{\tau}(b)}}{c_N} \int_{0}^{(\log N)^2}
\left(
    1 - \frac{N V_\tau^{(3)}(b)}{6 c_N^3} s^3+\frac{G'(b)s}{G(b)c_N}
\right)e^{-\frac{N V_\tau'(b)}{c_N} s 
    -\frac{N V_\tau''(b)}{2 c_N^2} s^2}\left(1 + {\rm O}\left(N \left(\frac{\log N}{c_N}\right)^4\right)\right)
\, ds.
    \end{align*}}
    The upper limit can be extended to $\infty$ with an exponentially small error and the resulting integral can be explicitly evaluated as
    \begin{align}\label{uj-out-int-ex}
        	\begin{aligned}
        	    &\sqrt{ \frac{2}{N V''_{\tau}(b)} }G(b)e^{ -N V_{\tau}(b) }f_{1}(\tilde{y}_{b})\bigg( 1 \\
	&\quad\left.+\sqrt{ \frac{2}{N V''_{\tau}(b)} }\left( \frac{V'''_{\tau}(b)}{6V''_{\tau}(b)}(2\tilde{y}_{b}^{3}+3\tilde{y}_{b})-\frac{G'(b)}{G(b)}\tilde{y}_{b}+\frac{1}{6f_{1}(\tilde{y}_{b}) }\left( \frac{3G'(b)}{G(b)}-\frac{V'''_{\tau}(b)}{V''_{\tau}(b)}(\tilde{y}_{b}^{2}+1) \right)  \right)  \right) .
        	\end{aligned}
    \end{align}
    where
    \begin{align*}
        \tilde{y}_b=\frac{\sqrt{N}V_{\tau}'(b)}{\sqrt{2V_{\tau}''(b)}}.
    \end{align*}
    The expansion \eqref{eq: uj asymb2} then follows from writing $V_{\tau}''(b)$ and $V_{\tau}'''(b) $ in terms of $\Delta q(b)$ using \eqref{deriV} and expanding the expression in large $N$ upon assuming $\sqrt{N}V_{\tau}'(b)$ to be ${\rm O}(1)$.
\end{proof}

\coro{}{For the case \(  0<t_\omega <\Delta_N \), the log ratio $\log\frac{u_j[a]}{u_j[0]}$ admits the $N\to\infty$  expansion
  \begin{equation}
  	\log\frac{u_j[a]}{u_j[0]}
	= a(\omega)-\varsigma_\omega\frac{a'(\omega)}{\sqrt N}\sqrt{\frac{2}{\Delta q(\omega)}}\;\frac{\tfrac12-x_\omega f_1(x_\omega)}{f_1(x_\omega)}+\mathrm{O}(N^{-1}).
\end{equation}
where $x_\omega$ and $f_1$ are given in \eqref{x-ome} and \eqref{eq:f1h1}.
}

Finally, for the case in which the critical point $r_{\tau}$ is outside and away from the hard walls, we have the following asymptotic expansions.
\propo{}{
If \( t_\omega \ge\Delta_N \), then the integral $u_j[a]$ admits the $N\to\infty$ expansion 
\begin{align}\label{eq: uj asymb1}
    	u_j[a] = \frac{e^{-N V_\tau(\omega)}\,\omega e^{a(\omega)}}{N\vert V_\tau'(\omega)\vert} \left( 1 \;+\; \frac{1}{N\,V_\tau'(\omega)} \left(\frac{1}{\omega}+a'(\omega)-\frac{V_\tau''(\omega)}{V_\tau'(\omega)}\right) + \mathrm{O}(V'_\tau(\omega)^{-4}N^{-2})\right).
\end{align}
}
\begin{proof}
    In the case $t_\omega \ge\Delta_N$, $V_{\tau}'(\omega)$ is large enough to dominate the exponent of the integrand in $u_j[a]$. Therefore, \eqref{eq: uj asymb1} can be obtained by expanding the expression with respect to $N$ with the assumption that $V_{\tau}'(\omega)$ is of order ${\rm O}(1)$.
\end{proof}

\coro{}{For the case \(  t_\omega \ge\Delta_N \), the log ratio of the integrals \eqref{uja}, $\log\frac{u_j[a]}{u_j[0]}$,  admits the $N\to\infty$  expansion
  \begin{equation}
  	\log\frac{u_j[a]}{u_j[0]}
	= a(\omega)+\frac{a'(\omega)}{NV_\tau '(\omega)}+\mathrm{O}(V'_\tau(\omega)^{-4}N^{-2}).
\end{equation}
}

\section{The ratio of partition functions and the Euler-Maclaurin formula}\label{S4}
Having derived the asymptotic expansions for $u_j[a]$, we are now able to assemble the summands $\log \frac{u_j[a]}{u_j[0]}$ by the Euler-Maclaurin formula and get the asymptotic expansions for the logarithmic averaged linear statistic $\log\left< e^{ \sum_{i}a(\lvert z_{i} \rvert ) } \right>_{\mathrm{a}}=\log (Z_{N}[q,a;\Omega]/Z_{N}[q,0;\Omega])$. As discussed in the last  paragraph of \S\ref{S2.2.2}, without loss of generality, we may fix the outer hard wall at $r=1$. It is easy to see that the quantities in Theorem \ref{thm1} do not change under the rescaling. So it suffices to prove the case when the hard walls are placed at $r=b,1$, which is the assumption of the previous section.

\begin{proof}[Proof of Theorem \ref{thm1}]
    Assuming the relation $r_0<b<1<r_1$ when both hard walls are inside the droplet, we split the sum as $S^{\rm out}_-+S^{\rm in}+S^{\rm out}_+$, where 
    \begin{align}
        & S^{\rm out}_- =\sum_{j=0}^{\lfloor N\tau_b\rfloor-1}\log\frac{u_j^{\mathrm a}[a]}{u_j^{\mathrm a}[0]},\, S^{\rm in}=\sum_{j=\lfloor N\tau_b\rfloor}^{\lfloor N\tau_0\rfloor-1}\log\frac{u_j^{\mathrm a}[a]}{u_j^{\mathrm a}[0]},\,S^{\rm out}_+ = \sum_{j=\lfloor N\tau_0\rfloor}^{N-1}\log\frac{u_j^{\mathrm a}[a]}{u_j^{\mathrm a}[0]}.
    \end{align}
    
    Combinations of these three sums, together with a particular choice of $\tau_b,\tau_0$, reproduce all the hard wall cases. The two-soft-wall case is obtained by setting $\tau_b=0,\,\tau_0=1$ in $S^{\rm in}$ and assuming the summand falls entirely in the bulk regime. The presence of a hard wall at the droplet boundary can be produced by letting $\tau_b=0$ or $\tau_0=1$ in $S^{\rm in} $ and including the edge regime for the summand at the corresponding end. The presence of a hard wall inside the droplet is given by the summation of $S^{\rm in}$ with one of the "outer" summations, in which case all the asymptotic regimes need to be considered. To be more specific, the complete proof requires consideration of the following cases
    \begin{itemize}
        \item Two soft walls ($b<r_0<r_1<1$)
        \begin{align*}
            \log\left\langle e^{\sum_{i}a(|z_i|)}\right\rangle_{\mathrm a}=\left(S^{\rm in,b}_0+S^{\rm in,b}_1\right) \bigg\vert_{\substack{\tau_b\to 0\\\tau_0\to 1}}
        \end{align*}
        \item One soft wall and one hard wall at droplet boundary ($b=r_0<r_1<1$ or $b<r_0<r_1=1$)
        \begin{align*}
            \log\left\langle e^{\sum_{i}a(|z_i|)}\right\rangle_{\mathrm a}=\left(S^{\rm in,b}_0+S^{\rm in,b}_1+S^{\rm in,e}_-\right) \bigg\vert_{\substack{\tau_b\to 0\\\tau_0\to 1}} \text{ or } \left(S^{\rm in,b}_0+S^{\rm in,b}_1+S^{\rm in,e}_+\right) \bigg\vert_{\substack{\tau_b\to 0\\\tau_0\to 1}}
        \end{align*}
         \item One soft wall and one hard wall inside the droplet ($b<r_0<1<r_1$ or $r_0<b<r_1<1$)
         {\small\begin{align*}
             \log\left\langle e^{\sum_{i}a(|z_i|)}\right\rangle_{\mathrm a}= S^{\rm out }_-+\left(S^{\rm in,b}_0+S^{\rm in,b}_1+S^{\rm in,e}_-\right) \bigg\vert_{\substack{\tau_0\to 1}} \text{ or } S^{\rm out }_++\left(S^{\rm in,b}_0+S^{\rm in,b}_1+S^{\rm in,e}_-\right) \bigg\vert_{\substack{\tau_b\to 0}}
         \end{align*}}
         \item Two hard walls at the droplet boundary ($b=r_0<r_1=1$)
         \begin{align*}
            \log\left\langle e^{\sum_{i}a(|z_i|)}\right\rangle_{\mathrm a}=S^{\rm in}\bigg\vert_{\substack{\tau_b\to 0\\\tau_0\to 1}}
         \end{align*}
         \item One hard wall at droplet boundary, one hard wall inside ($b=r_0<1<r_1$ or $r_0<b<r_1=1$)
         \begin{align*}
             \log\left\langle e^{\sum_{i}a(|z_i|)}\right\rangle_{\mathrm a}=S^{\rm in}\bigg\vert_{\tau_0\to 1}+S^{\rm out}_- \quad  \text{ or } \quad S^{\rm in}\bigg\vert_{\tau_b\to 0}+S^{\rm out}_+
         \end{align*}
         \item Two hard walls inside the droplet ($r_0<b<1<r_1$)
         \begin{align*}
             \log\left\langle e^{\sum_{i}a(|z_i|)}\right\rangle_{\mathrm a}=S^{\rm out }_-+S^{\rm in}+S^{\rm out }_+
         \end{align*}
         \item The non-overlapping case I ($r_0=1$ or $r_1=b$)
         \begin{align*}
             S_-^{\rm out}\bigg\vert_{\tau_b\to 1} \quad  \text{ or } \quad S_+^{\rm out} \bigg\vert_{\tau_0\to 0}
         \end{align*}
         \item The non-overlapping case II ($r_0>1$ or $r_1<b$)
         \begin{align*}
             \log\left\langle e^{\sum_{i}a(|z_i|)}\right\rangle_{\mathrm a}=S^{\rm out,b}_{-,0}\bigg\vert_{\tau_b\to 1}+S^{\rm out,b}_{-,1}\bigg\vert_{\tau_b-\Delta_N\to 1} \quad  \text{ or } \quad S^{\rm out,b}_{+,0}\bigg\vert_{\tau_0\to 0}+S^{\rm out,b}_{+,1}\bigg\vert_{\tau_0+\Delta_N\to 0}
         \end{align*}
    \end{itemize}
    Here the split sums $S^{\rm in,b}_0,\,S^{\rm in,b}_1,\,S^{\rm in,e}_-,\,S^{\rm in,e}_+ $ are defined in \eqref{Sin-split}; the outer split sums $S^{\rm out,b}_{-,0},S^{\rm out,b}_{-,1}$ are defined in \eqref{Sout-split} with $S^{\rm out,b}_{+,0},S^{\rm out,b}_{+,1}$ defined in a similar manner. The proof is then completed through combinations of \eqref{Sinb0}, \eqref{Sinb1}, \eqref{Sine-}, \eqref{Sout-asy}, \eqref{Soutb0} and \eqref{Soutb1} below.
\end{proof}

\propo{}{
The $N\to \infty$ asymptotic expansion up to ${\rm O}(1)$ of $S^{\rm in} $ is given by
\begin{align}\label{Sin-asy}
    \begin{aligned}
        S^{\rm in}=	&N\int_{\tau_b}^{\tau_0}a(r_\tau)\,d\tau+\eta_N^b a(b)-\eta_N^1 a(1)+\frac14(2\log 2-1)\bigl(ba'(b)-a'(1)\bigr)\\
	&+\frac14\int_b^1 r\,a'(r)\Bigl(a'(r)-\frac{\partial_r\Delta q(r)}{\Delta q(r)}\Bigr)dr+\mathrm o(1),
    \end{aligned}
\end{align}
where $\eta_N^b=\{N\tau_b\},\,\eta_N^1=\{N\tau_0\}$.
}
\begin{proof}
    According to different asymptotic regimes, $S^{\rm in} $ can be subdivided into
    \begin{align}\label{Sin-split}
        \begin{aligned}
            S^{\rm in}=&\sum_{j=\lfloor N \tau_b \rfloor}^{\lfloor N\tau_0 \rfloor-1}a(r_\tau)+\sum_{j=\lfloor N(\tau_b+\delta_N)\rfloor}^{\lfloor N(\tau_0-\delta_N)\rfloor}\frac{1}{N}\left(\frac{a''(r_\tau)+a'(r_\tau)^2}{2\Delta q(r_\tau)}+\frac{3 a'(r_\tau)}{2r \Delta q(r_\tau)}-\frac{a'(r_\tau)\partial_r\Delta q(r_\tau)}{2 \Delta q(r_\tau)^2}\right)\\
        &-\sum_{j=\lfloor N\tau_b\rfloor}^{\lfloor N(\tau_b+\delta_N)\rfloor-1}ba'(b)\frac{e^{-x_b^2}}{2\sqrt{\pi}f_2(x_b)}\frac{dx_b}{dj}-\sum_{j=\lfloor N(\tau_0-\delta_N)\rfloor+1}^{\lfloor N\tau_0\rfloor-1}a'(1)\frac{e^{-x_1^2}}{2\sqrt{\pi}f_2(x_1)}\frac{dx_1}{dj}\\
        =&S^{\rm in,b}_0+S^{\rm in,b}_1+S^{\rm in,e}_-+S^{\rm in,e}_+ .
        \end{aligned}
    \end{align}
    Making use of the Euler-Maclaurin formula for the first sum, we get
    \begin{align}\label{Sinb0}
        &\sum_{j=\lfloor N\tau_b\rfloor}^{\lfloor N\tau_0\rfloor-1} a(r_\tau)
	= N\int_{\tau_b}^{\tau_0}a(r_\tau)\,d\tau
	+ a(b)\bigl(\{N\tau_b\}+\tfrac12\bigr)-a(1)\bigl(\{N\tau_0\}+\tfrac12\bigr)+\mathrm o(1).
    \end{align}
    Through the change of variable $d\tau =\frac{r\Delta q(r)}{2}dr$, the second sum contributes
    \begin{align}\label{Sinb1}
        \begin{aligned}
            S^{\rm in,b }_1&=\frac{1}{4}\int_{b}^1 \left( r(a''(r)+a'(r)^{2})+3a'(r)-\frac{a'(r)r\partial_{r}\Delta q(r)}{\Delta q(r)} \right)dr+{\rm o}(1)\\
        &=\frac14\bigl(a'(1)-b\,a'(b)\bigr)+\frac12\bigl(a(1)-a(b)\bigr)+\frac14\int_b^1 r\,a'\Bigl(a'-\frac{\partial_r\Delta q}{\Delta q}\Bigr)dr+{\rm o}(1).
        \end{aligned}
    \end{align}
    For the third sum, it is enough to consider the leading integral in the Euler-Maclaurin formula, and by extending the lower limit to $-\infty$ we get
    \begin{align}\label{Sine-}
        \begin{aligned}
            -\sum_{j=\lfloor N\tau_b\rfloor}^{\lfloor N(\tau_b+\delta_N)\rfloor-1}ba'(b)\frac{e^{-x_b^2}}{2\sqrt{\pi}f_2(x_b)}\frac{dx_b}{dj}&=ba'(b)\int^{0}_{-\frac{1}{b}\sqrt{ \frac{2N}{\Delta q(b)} }\delta_{N}}\frac{e^{ -x^{2} }}{2\sqrt{ \pi }f_{2}(x)}dx+{\mathrm{ o}}(1)\\
        &=\frac{ba'(b)}{2}\log 2+{\mathrm{o}}(1).
        \end{aligned}
    \end{align}
    A similar calculation shows that the contribution of the last sum is $-\frac{a'(1)}{2}\log2$. Therefore, by adding up the above formulas, we get equation \eqref{Sin-asy}.
\end{proof}

\rem{}{
It can be seen from the above calculations that the bulk correction produces a bulk integral $\frac14\int_b^1 r\,a'\Bigl(a'-\frac{\partial_r\Delta q}{\Delta q}\Bigr)dr$, and an edge term $\pm\frac{1}{4}\omega_{\pm} a'(\omega_{\pm}) $ for each endpoint (the term $\frac{1}{2}(a(1)-a(b)) $ cancels with the one in the first sum). If a hard wall is imposed at the droplet boundary, the analysis of the edge summations shows an extra contribution $\mp \frac{\omega_{\pm}a'(\omega_{\pm})}{2}\log 2$. These are in line with the formulas of $\mathcal{E}_{\pm}[a]$ in Theorem \ref{thm1} for the soft-wall case and the case when a hard wall is placed at the droplet boundary.
}

\propo{}{
The $N\to \infty$ asymptotic expansions up to ${\rm O}(1)$ of $S^{\rm out}_- $ and $S^{\rm out}_+ $ are given by
\begin{align}\label{Sout-asy}
    \begin{aligned}
        &S^{\rm out}_-=a(b)\bigl(N\tau_b-\{N\tau_b\}\bigr)
	+\frac{b\,a'(b)}{4}\Bigl(\log N+2\log\frac{\tau_b}{b}+\log\frac{2\pi}{\Delta q(b)}\Bigr)+\mathrm o(1),\\
    &S^{\rm out}_+=a(1)\bigl(N(1-\tau_0)+\{N\tau_0\}\bigr)
	-\frac{a'(1)}{4}\Bigl(\log N+2\log(1-\tau_0)+\log\frac{2\pi}{\Delta q(1)}\Bigr)+\mathrm o(1).
    \end{aligned}
\end{align}
}
\begin{proof}
    Here we only prove the formula for $S^{\rm out}_- $; the one for $S^{\rm out}_+  $ can be proved in a similar manner. We first split the sum into
    \begin{align}\label{Sout-split}
        \begin{aligned}
            S^{\rm out}_-&=\sum_{j=0}^{\lfloor N\tau_b\rfloor-1}a(b)+\sum_{j=0}^{\lfloor N(\tau_b-\Delta_N)\rfloor-1}\frac{a'(b)}{NV_\tau'(b)}-\sum_{j=\lfloor N(\tau_b-\Delta_N)\rfloor}^{\lfloor N\tau_b\rfloor-1}ba'(b)\frac{1-2x_b f(x_b)}{2f_1(x_b)}\frac{dx_b}{dj}\\
            &:=S^{\rm out,b}_{-,0}+S^{\rm out,b}_{-,1}+S^{\rm out, e}_- ,
        \end{aligned}
    \end{align}
    where $\frac{dx_b}{dj}=-\frac{1}{b}\sqrt{\frac{2}{N \Delta q(b)}}$.
    The leading term is just a sum of constants, which gives 
    \begin{align}\label{Soutb0}
        \sum_{j=0}^{\lfloor N\tau_b\rfloor-1}a(b)=a(b)(N\tau_b-\{N\tau_b\}).
    \end{align}
    An analysis of the second sum shows that 
    \begin{align}\label{Soutb1}
        \begin{aligned}
            \sum_{j=0}^{\lfloor N(\tau_b-\Delta_N)\rfloor-1}\frac{a'(b)}{NV_\tau'(b)}&=-\sum_{j=0}^{\lfloor N(\tau_b-\Delta_N)\rfloor-1}\frac{ba'(b)}{2N(\tau-\tau_b)}=-\frac{ba'(b)}{2}\int_0^{\tau_b-\Delta_N}\frac{d\tau}{\tau-\tau_b}+{\rm o}(1)\\
        &=-\frac{ba'(b)}{2}(\log \Delta_N-\log \tau_b)+{\rm o}(1).
        \end{aligned}
    \end{align}
    Finally, the edge contribution from the outer side is
    \begin{align}
        -\sum_{j=\lfloor N(\tau_b-\Delta_N)\rfloor}^{\lfloor N\tau_b\rfloor-1}ba'(b)\frac{1-2x_b f(x_b)}{2f_1(x_b)}\frac{dx_b}{dj}=ba'(b)\int_{0}^{\frac{1}{b}\sqrt{\frac{2N}{\Delta q(b)}}\Delta_N}\frac{1-2xf_1(x)}{2f_1(x)}dx+{\rm o}(1).
    \end{align}
    Recall that by definition, $\sqrt{N}\Delta_N\to \infty$ as $N\to \infty$, so we can extend the upper limit to $\infty$ after subtracting $\frac{1}{2x+1}$ to make the integral convergent. Therefore, we have
    \begin{align*}
        ba'(b)\int_{0}^{\frac{1}{b}\sqrt{\frac{2N}{\Delta q(b)}}\Delta_N}\frac{1-2xf_1(x)}{2f_1(x)}dx&=ba'(b)\int_{0}^\infty \left(\frac{1}{2f_1(x)}-x-\frac{1}{2x+1}\right)dx\\
        &\qquad+ba'(b)\int_{0}^{\frac{1}{b}\sqrt{\frac{2N}{\Delta q(b)}}\Delta_N}\frac{dx}{2x+1}+{\rm o}(1)\\
        =\frac{ba'(b)}{4}&\left(\log \pi +\log \frac{2N}{b^2\Delta q(b)}\right)+\frac{ba'(b)}{2}\log \Delta_N+{\rm o}(1).
    \end{align*}
    Equation \eqref{Sout-asy} is then proved by combining the above formulas.
\end{proof}

\rem{}{
The above calculation indicates a contribution
\begin{align}
    \mp \frac{\omega_{\pm}a'(\omega_{\pm})}{4}\left(2 \log m_\pm+ \log \pi +\log \frac{2N}{\omega_\pm^2\Delta q(\omega_\pm)}\right) 
\end{align}
from the outer part when a hard wall is placed inside the droplet. It's also worth noticing that the ${\rm O}(\log N)$ term depends on both $a'(r)$ and the end points, which makes the cylinder geometry rather special for having no $\log N$ term in the asymptotic expansion of $\log Z_N^{\rm c}(q)$.
}

\coro{C4.1}{
The $N\to \infty$ asymptotic expansion of the logarithmic ratio $\log {{Z}_N^{\rm c}(q) \over {Z}_N^{\rm a}(q)}$ is given by
\begin{align}\label{asy-Zc/Za}
    \log {{Z}_N^{\rm c}(q) \over {Z}_N^{\rm a}(q)} = N\left(\frac{1}{2}(q(\omega_+)-q(\omega_-))-\log \omega_{+} \right)+\frac{1}{4}\log \frac{\omega_+\Delta q(\omega_+)}{\omega_-\Delta q(\omega_-)}+\mathcal{E}_{+}+\mathcal{E}_{-}+{\rm o}(1),
\end{align}
where
\begin{align}
    \mathcal{E}_{\pm}=
\begin{cases}
    \mp \frac{1}{4},\quad \text{soft edge},\\
    \pm \frac{1}{4}\Bigl(\log N + 2\log\dfrac{m_{\pm}}{\omega_{\pm}} + \log\dfrac{8\pi}{\Delta q(\omega_{\pm})} - 1\Bigr),\quad \text{hard wall strictly inside},\\
    \pm\frac{1}{4}\bigl(2\log 2-1\bigr),\quad \text{hard wall at the droplet boundary}.
\end{cases}
\end{align}
}

\section{ 
$Z_N^{\rm c}(Q^{\rm c})$ with $Q^{\rm c}(Y) = { W \over L} Y^2$ and
$Z_N^{\rm c}(q)$ with $q=r^{2 \mathfrak b}$}\label{S5}

\subsection{Proof of Proposition \ref{P1.1}}\label{S5.1}
The large $N$ expansion of $Z_N^{\rm c}(Q^{\rm c})$ with $Q^{\rm c}(Y) = { W \over L } Y^2$ in the case of soft walls, or with hard wall(s) positioned at the droplet boundary, is specified in Proposition \ref{P1.1}. 
This can be established by specialising Theorem 
\ref{thm2hw} according to \eqref{eq6}, \eqref{eq7}.
This specialisation together with the requirement that any hard wall be at the droplet boundary implies
\begin{equation}\label{b.3x}
\tau_- = 0, \quad \tau_+ = 1, \quad \omega_- = r_0, \quad 
 \omega_+ = r_1,
\end{equation}
as follows from \eqref{ome+-} and \eqref{m-pm}. We can then compute from \eqref{I-muq}, \eqref{E-muq}, 
\eqref{eta} and \eqref{F1} 
that 
\begin{equation}\label{b.3y}
I_{\mathbb{S}\cap\mathbb{A}}[\mu_q^{\Omega}]=
- {\pi L \over 6 W}
, \quad
E_{\mathbb{S}\cap\mathbb{A}}^{\rm bulk}[\mu_q^{\Omega}] =
\log \Big ( {2L \over W} \Big ), \quad \eta = 0, \quad
F_{\mathbb{S}\cap \mathbb{A}}[q] =  {\pi L \over 6 W}.
\end{equation}
Specialisation of \eqref{3.3} when used in the soft wall,
and hard wall at an edge formulas for 
$\mathcal{E}_{\pm}^{(N)}$ in Theorem 
\ref{thm2hw}, gives that
\begin{equation}\label{b.3z}
\mathcal{E}_{-}^{(N)} +
\mathcal{E}_{+}^{(N)} = 0.
\end{equation}
It remains to consider $\mathcal{E}_{\pm}^{(1)}$ in
Theorem  
\ref{thm2hw}. Noting from \eqref{constin} that
$\tilde{\alpha}^{\rm in} + 2 \beta^{\rm in} = {1 \over 2} \log 2$ we see that for the present specialisation, for two soft walls, or a single hard wall at the edge, or two hard walls at the edge
\begin{equation}
\mathcal{E}_{-}^{(1)} +
\mathcal{E}_{+}^{(1)} = 0.
\end{equation}
From this working, \eqref{b.4}--\eqref{b.6} follow.

\subsection{The Mittag-Leffler case $Z_N[r^{2\mathfrak{b}},2\alpha\log r;\Omega]$}\label{S5.2}
In \cite{Ch24}, the author considered the random normal matrix model with the joint eigenvalue p.d.f proportional to \eqref{1.9} with $\beta=2$ and the potential $Q(z)=|z|^{2\mathfrak{b}}+\frac{2\alpha}{N}\log |z| $. The one-body terms are said to give rise to the Mittag-Leffler case. The partition function for that ensemble with two hard wall constraints, in our notation, corresponds to $Z_N^{\rm a}[r^{2\mathfrak{b}}+(2\alpha/N)\log r;\Omega]$. With $\alpha=-1/2$, this reduces to the partition function for cylinder geometry with $q(r)=r^{2\mathfrak{b}}$, and so  provides an opportunity for a consistency check on our results. In doing so we are making use of the fact that
Theorem \ref{thm2hw} remains valid in the case that the droplet is a disk rather than an annulus (the former being the support in the Mittag-Leffler case), provided that there is a lower hard wall; recall Remark \ref{R1.2}.

Let us then specialize 
Theorem \ref{thm2hw}
to the Mittag-Leffler case. Suppose there are two effective hard walls at $r=\rho_-,\rho_+$, with $0<\rho_{-}<\rho_{+}<\mathfrak b^{-1/(2\mathfrak b)}$. Then we have
\begin{align}
    \Omega=[\rho_-,\rho_+], \quad \omega_\pm=\rho_\pm,\quad \tau_\pm=\mathfrak{b}\rho_\pm^{2\mathfrak{b}},\quad \Delta q(r)=4\mathfrak{b}^2r^{2\mathfrak{b}-2}.
\end{align}
According to the $g=2$ case of Theorem 1.9 in \cite{Ch24}, and after adding the soft wall partition function known from \cite{BKS23},
\cite{AFLS25}, the asymptotic expansion of $\log Z_N^{\rm a}[r^{2\mathfrak{b}}+(2\alpha/N)\log r;\Omega]$ is given by
\begin{align}
    C_{1}N^{2}+C_{2}N\log N+C_{3}N+C_{4}\sqrt{ N }+C_{6}+\mathrm{O}(N^{-1/12}),
\end{align}
where the coefficients are given by
\begin{align}\label{Charlier-Cs}
	\begin{aligned}
	    & C_{1}=\frac{\mathfrak{b}}{4}(\rho_{+}^{4\mathfrak{b}}-\rho_{-}^{4\mathfrak{b}})-\rho_{+}^{2\mathfrak{b}}+\log \rho_{+}, \\
	& C_{2}=\frac{\mathfrak{b}}{2}(\rho_{+}^{2\mathfrak{b}}-\rho_{-}^{2\mathfrak{b}})-1, \\
	& C_{3}=\left( \frac{1}{2}+\frac{2\alpha+1}{2\mathfrak{b}}-\frac{1}{2}\log \frac{2\pi}{\mathfrak{b}} \right)(\tau_{-}-\tau_{+})-m_{-}\log m_{-}-m_{+}\log m_{+} \\
	& \qquad+\frac{1}{2}(\tau_{-}\log \tau_{-}-\tau_{+}\log \tau_{+})+(1+2\alpha)\log \rho_{+}+1+\log \pi,  \\
	& C_{4}=\left( \rho_{+}\sqrt{ \Delta q(\rho_{+}) }+\rho_{-}\sqrt{ \Delta q(\rho_{-}) } \right) (\gamma^{\mathrm{in}}+\gamma^{\mathrm{out}}), \\
	& C_{6}=\frac{1}{2}(1+2\alpha)(\log \tau_{-}-\log(1-\tau_{+}))+\frac{\rho_{+}^{2}\Delta q(\rho_{+})}{4(1-\tau_{+})}+\mathfrak{b}+\frac{\mathfrak{b}^{2}+6\alpha^{2}+6\alpha+1}{6}\log \frac{\rho_{+}}{\rho_{-}}.
	\end{aligned}
\end{align}
On the other hand, specialising Theorem \ref{thm2hw} with $q=r^{2\mathfrak{b}}$ and taking into account the factor $(2\pi )^{-N}$ shows
\begin{align}
    	\log Z_{N}^{\mathrm{c}}[q;(\rho_{-},\rho_{+})]=\hat{C}_{1}N^{2}+\hat{C}_{2}N\log N+\hat{C}_{3}N+\hat{C}_{4}\sqrt{ N }+\hat{C}_{6}+\mathrm{o}(1),
\end{align}
where
\begin{align}\label{our-Cs}
    \begin{aligned}
		& \hat{C}_{1}=-I_{\mathbb{S}\cap \mathbb{A}}[\mu_{q}^{\Omega}], \quad \hat{C}_{2}=-\frac{1}{2}(1+\eta),\\
		& \hat{C}_{3}=-\frac{1}{2}\left( E_{\mathbb{S}\cap \mathbb{A}}^{\rm bulk}[\mu_{q}^{\Omega}]+(1+\eta)\log 2+\mathcal{E}_{-}^{(N)}+\mathcal{E}_{+}^{(N)} \right)+\log(2\pi),\\
		& \hat{C}_{4}=\left( \rho_{+}\sqrt{ \Delta q(\rho_{+}) }+\rho_{-}\sqrt{ \Delta q(\rho_{-}) } \right) (\gamma^{\mathrm{in}}+\gamma^{\mathrm{out}}),\\
		& \hat{C}_{6}=F_{\mathbb{S}\cap \mathbb{A}}[q]+\mathcal{E}_{-}^{(1)}+\mathcal{E}_{+}^{(1)}.
	\end{aligned}
\end{align}
The required quantities are evaluated as:
\begin{align}
    \begin{aligned}
        & I_{\mathbb{S}\cap \mathbb{A}}\left[\mu_q^{\Omega}\right]=  \tau_{-} \rho_{-}^{2 \mathfrak{b}}-\tau_{-}^2 \log \rho_{-}+\left(1-\tau_{+}\right) \rho_{+}^{2 \mathfrak{b}}-\left(1-\tau_{+}^2\right) \log \rho_{+} \\
        & \qquad \qquad \quad +\int_{\tau_{-}}^{\tau_{+}}\left(r_\tau^{2 \mathfrak{b}}-2 \tau  \log r_\tau\right) d \tau \\
        & \qquad \qquad= \frac{1}{4 \mathfrak{b}}\left(\tau_{-}^2-\tau_{+}^2\right)+\rho_{+}^{2 \mathfrak{b}}-\log \rho_{+},\\
        & E_{\mathbb{S}\cap \mathbb{A}}^{\rm bulk}[\mu_{q}^{\Omega}]=\left(\tau_{+}-\tau_{-}\right) \log \frac{\mathfrak{b}^2}{\pi}+\left(1-\frac{1}{\mathfrak{b}}\right)\left[\tau_{+} \log \omega_{+}^{2 \mathfrak{b}}-\tau_{-} \log \omega_{-}^{2 \mathfrak{b}}-\tau_{+}+\tau_{-}\right],\\
        &F_{\mathbb{S}\cap \mathbb{A}}[r^{2\mathfrak{b}}]=\left( \frac{1}{6}(\mathfrak{b}-1)^{2}-\frac{\mathfrak{b}}{6} \right) \log \frac{\rho_{+}}{\rho_{-}},\\
        &\mathcal{E}_{-}^{(1)}+\mathcal{E}_{+}^{(1)}=\frac{1}{4}\log \frac{\rho_{+}^{2\mathfrak{b}-1}}{\rho_{-}^{2\mathfrak{b}-1}}+\left( \frac{\mathfrak{b}^{2}\rho_{+}^{2\mathfrak{b}}}{m_{+}}+\frac{\mathfrak{b}^{2}\rho_{-}^{2\mathfrak{b}}}{m_{-}} \right)\\
		&\qquad \qquad \quad =\frac{2\mathfrak{b}-1}{4}\log \frac{\rho_{+}}{\rho_{-}}+\frac{\mathfrak{b}^{2}\rho_{+}^{2\mathfrak{b}}}{m_{+}}+\mathfrak{b}.
    \end{aligned}
\end{align}
Substituting this into \eqref{our-Cs}, we see that $\hat{C}_1,\dots,\hat{C}_6$ matches exactly with $C_1,\dots,C_6$ in \eqref{Charlier-Cs} with $\alpha=-1/2$.
The remaining $\alpha$-dependent part
\begin{align*}
(2 \alpha+1) \left(\frac{1}{2 \mathfrak{b}}\left(\tau_{-}-\tau_{+}\right)+\log \rho_{+}\right)N+\frac{1}{2}(1+2 \alpha)\log \frac{\tau_{-}}{1-\tau_{+}}+\left(\alpha+\frac{1}{2}\right)^2 \log \frac{\rho_{+}}{\rho_{-}} 
\end{align*}
also matches the results in Theorem \ref{thm1} and Corollary \ref{C4.1}.


\section*{Acknowledgements}
The work of PJF is supported by a grant from the Australian Research Council, Discovery Project
DP250102552. 
 BJS is supported by  the Shanghai Jiao Tong
University Overseas Joint Postdoctoral Fellowship Program.
Participation in the beginning stages of this project by Matthias Allard and Sampad Lahiry is acknowledged. We appreciate too the helpful feedback on a draft of this work from Sung-Soo Byun and Christophe Charlier.

\appendix
\section{Appendix: Poisson equation for cylinder geometry}\label{appendix0}

In two-dimensional electrostatics
 the interaction of each pair of particles, at $\vec{r}$ and $\vec{r}\, '$ say, is via a pair potential $\Psi(\vec{r},\vec{r}\, ')$ determined as the solution of the Poisson equation with delta function source
\begin{equation}\label{0.2}
\nabla^2 \Psi(\vec{r},\vec{r}\, ') = - 2 \pi \delta(\vec{r}- \vec{r} \,').
\end{equation}
Actually, a further assumption is required, namely that the two-dimensional domain be of infinite extent (the plane or the surface of a cylinder); for a compact surface such as the sphere or topological torus, the right hand side of (\ref{0.2}) must be modified by the addition of an appropriate constant corresponding to a charge neutrality constraint.

Here we consider the specialisation of (\ref{0.2}) to the surface of a cylinder. Take the
 axis of the cylinder to be along the $z$-axis, let the radius to the boundary equal $a$, and specify a point on the boundary using the azimuthal angle $\phi$ and its height $z$. The Poisson equation with a delta function source in this setting reads
\begin{equation}\label{0.4}
\bigg ( {\partial^2 \over \partial z^2} + {1 \over a^2}
{\partial^2 \over \partial \phi^2} \bigg )
\Psi((z,\phi),(z',\phi')) = - {2 \pi \over a}
\delta(z-z') \delta(\phi-\phi').
\end{equation}
This equation
is to be solved subject to the requirement that it is periodic, period $2 \pi$, in $\phi$ (and also in $\phi'$ since $\Psi$ is symmetric with respect to interchange of
$(z,\phi)$ and $(z',\phi')$). Setting $a \phi = x$ so that $0 \le x \le W$, where $W = 2 \pi a$ is the circumference length of the boundary, and setting $z = y$
(similarly too for $\phi'$ and $y'$)
shows that (\ref{0.4}) can be rewritten
\begin{equation}\label{0.5}
\bigg ( {\partial^2 \over \partial x^2} +
{\partial^2 \over \partial y^2} \bigg )
\Psi((x,y),(x',y'))
= - 2 \pi\delta(x-x') \delta(y-y').
\end{equation}
This is identical to the planar Poisson equation (\ref{0.2}),
except that we require the semi-periodic boundary condition
$\Psi((x,y),(x',y')) = \Psi((x+W,y),(x',y'))$. The solution
is readily verified to be
\begin{equation}\label{0.6}
\Psi((x,y),(x',y')) = - \log \bigg ( \sin\Big ( \pi (x - x' + i (y - y'))/W \Big ) \Big ( {W \over \pi} \Big ) \bigg );
\end{equation}
see e.g.~\cite[Eq.~(2.2)]{CFTW15}.
Here the factor $W/\pi$ outside of the sine function has been chosen so that for $W \to \infty$, the planar solution is obtained.

\section{Appendix: Euler-Maclaurin formula}\label{appendixA}
Essential to our analysis is that for the class of sums encountered, the large $N$ asymptotic expansion can be deduced by applying the Euler-Maclaurin summation formula \cite{DLMF}, which we record here for reference.

\propo{prop: EM formula}{(Euler-Maclaurin formula) Let $f(x)$ be $2k$ times differentiable  on the interval $[p,q]$. Denote by $\{B_{2k} \}$ the even indexed Bernoulli numbers ($B_2=\frac{1 }{ 6}, B_4 = - \frac{1 }{ 30}, \dots$)
\Eq{EM1}{
\sum _{i=p+1}^{q-1}f(i)=\int _{p}^{q}f(x)~{\rm {d}}x-{\frac {f(p)+f(q)}{2}}+\sum _{j=1}^{k}{\frac {B_{2j}}{(2j)!}}\left(f^{(2j-1)}(q)-f^{(2j-1)}(p)\right)+R_{2k}
}
\Eq{EM2}{
\sum _{i=p}^{q}f(i)=\int _{p}^{q}f(x)~{\rm {d}}x+{\frac {f(p)+f(q)}{2}}+\sum _{j=1}^{k}{\frac {B_{2j}}{(2j)!}}\left(f^{(2j-1)}(q)-f^{(2j-1)}(p)\right)+R_{2k},
}
where the remainder $R_{2k}$ satisfies the bound
$|R_{2k}| \le c_{2k}
\int_p^q | f^{(2k)}(x) | \, dx$, with $c_{2k} = 2 \zeta(2k) / (2 \pi)^{2k}$.
}

\section{Appendix: Annulus geometry and two hard walls}\label{appendixB}
In this appendix, using direct working we are going to derive the asymptotic expansion of $\log {Z}_N^{\rm c}$ with two hard walls up to and including the constant term. We begin by introducing some quantities. Let $\mathbb{A}$ be the hard wall constraint given in \eqref{mathbbA}. In potential theory, one often considers the weighted logarithmic energy
\begin{align}
    I_{\mathbb{S}\cap\mathbb{A}}[\nu]
:=\iint_{\mathbb{A}}
\log\frac{1}{|z-w|}\,d\nu(z)d\nu(w)
+\int_{\mathbb{A}}q(|z|)\,d\nu(z).
\end{align}
This is a functional such that the equilibrium measure \eqref{muq-ome} is its unique minimizer  among all probability measures supported on $\mathbb{A}$. Let $R(\tau)$ be the minimum point of $V_\tau(r)$
as specified by (\ref{3.3})
in $\Omega$ given by
\begin{align}
    R(\tau)=
\begin{cases}
\omega_{-}, & 0\leq\tau\leq\tau_-,\\
r_\tau, & \tau_-\leq\tau\leq\tau_+,\\
\omega_{+}, & \tau_+\leq\tau\leq1.
\end{cases}
\end{align}
Then for every radial test function $f$, one has
$$
\int_{\mathbb C}f(|z|)\,d\mu_q^\Omega(z)
=\int_0^1f(R(\tau))\,d\tau.
$$
Making use of the angular average identity
\begin{align}
    \frac{1}{(2\pi)^2}\int_0^{2\pi}\int_0^{2\pi}
\log|re^{i\theta}-se^{i\phi}|\,d\theta d\phi
=\log\max\{r,s\},
\end{align}
we see that the first term in the weighted logarithmic energy $I_{\mathbb{S}\cap\mathbb{A}}[\mu_q^{\Omega}]$ can be rewritten as
\begin{align}
    \iint\log|z-w|\,d\mu_q^\Omega(z)d\mu_q^\Omega(w)
=2\int_0^1\tau\log R(\tau)\,d\tau.
\end{align}
Combined with the second term, we get the simple form
\begin{align}
    I_{\mathbb{S}\cap\mathbb{A}}[\mu_q^{\Omega}]=\int_0^1V_\tau(R(\tau))\,d\tau.
\end{align}
Splitting the integral into bulk and edge contributions shows
\begin{align}\label{I-muq}
    \begin{aligned}
I_{\mathbb{S}\cap\mathbb{A}}[\mu_q^{\Omega}]={}&\int_0^{\tau_-}V_\tau(\omega_{-})\,d\tau
+\int_{\tau_-}^{\tau_+}V_\tau(r_\tau)\,d\tau
+\int_{\tau_+}^{1}V_\tau(\omega_{+})\,d\tau\\
={}&\tau_-q(\omega_{-})-\tau_-^2\log\omega_{-}\\
&+\frac12\int_{\omega_{-}}^{\omega_{+}}
r\Delta q(r)\left[q(r)-rq'(r)\log r\right]dr\\
&+(1-\tau_+)q(\omega_{+})
-(1-\tau_+^2)\log\omega_{+}.
\end{aligned}
\end{align}
For later use, denote $I_{\mathbb{S}\cap\mathbb{A}}^{\rm bulk}[\mu_q^{\Omega}]=\int_{\tau_-}^{\tau_+}V_\tau(r_\tau)\,d\tau$.
Also required is the bulk entropy
\begin{align}\label{E-muq}
    \begin{aligned}
        E_{\mathbb{S}\cap\mathbb{A}}^{\rm bulk}[\mu_q^{\Omega}]&:=\int_{\{\omega_{-}<|z|<\omega_{+}\}}
\log\left(\frac{\Delta q(|z|)}{4\pi}\right)
\frac{\Delta q(|z|)}{4}\,dA(z)\\
&=\frac12\int_{\omega_{-}}^{\omega_{+}}
r\Delta q(r)
\log\left(\frac{\Delta q(r)}{4\pi}\right)dr\\
&=\int_{\tau_-}^{\tau_+}
\log\left(\frac{\Delta q(r_\tau)}{4\pi}\right)d\tau.
    \end{aligned}
\end{align}
In what follows, we introduce some universal ($q(r)$ independent) integrals appearing in the expansions, which in their integrands typically involve
\Eq{eq: def phi}{
\Phi(x):=\log\left(\frac{1}{2}\erfc(x)\right),
\quad \Phi'(x)=-\frac{2 e^{-x^2}}{\sqrt{\pi } \erfc(x)
},
}
where $\erfc$ is the usual complementary error function.
Thus we set
\Eq{constin}{
\tilde{\alpha}^{\rm in}&:=\int_{-\infty}^0  \frac{ \left(2x^2-1\right)}{6 }\Phi'(x) dx\approx 0.02284\\
\beta^{\rm in}&:=-\int_{-\infty}^0  \frac{\left(x^2+1\right) }{6}\Phi'(x) dx\approx 0.16186\\
\gamma^{\rm in}&:=-\frac{1}{\sqrt{2}}\int_{-\infty}^0 \Phi(x) dx\approx 0.23876,
}
and 
\Eq{constout}{
\tilde{\alpha}^{\rm out}&:=\frac{1}{6}\int_0^{\infty}\left(  -(2x^2-1)\Phi'(x)-4x^3+\frac{3}{x+1}\right)dx\approx 0.00866\\
\beta^{\rm out}&:=-\frac{1}{6}\int_0^{\infty}\lrbra{x(2x^2+3)+(x^2+1)\Phi'(x)}dx\approx 0.14742\\
\gamma^{\rm out}&:=-\frac{1}{\sqrt{2}}\int_0^{\infty} \log\left( \sqrt{\pi} x e^{x^2} \erfc(x) \right) dx\approx 0.91194.
}
Note that $\gamma^{\rm in}, \gamma^{\rm out}$ were introduced earlier in (\ref{gi}); the reference \cite{AFLS25} contains the same definite integrals, except that here we have modified the first in each list (and thus the reason for the tilde). It relates to the constants in \cite[eq.(2.4),(2.5)]{AFLS25} through
\begin{align}
    \tilde{\alpha}^{\rm in}=\alpha^{\rm in}-\frac{1}{2}\log 2,\quad \tilde{\alpha}^{\rm out}=-\alpha^{\rm out}+\frac{1}{4}\log \pi.
\end{align}
To simplify the notations, we introduce operators
\begin{align}\label{tpm}
    \mathcal{T}_{1,b}^+:f(x)\mapsto f(1)+f(b),\quad \tm :f(x)\to f(1)-f(b),
\end{align}
which will be used throughout the appendix.
We also define the total mass (see \eqref{m-pm}) of surface charges in the equilibrium measure induced by the hard walls
\begin{align}\label{eta}
    \eta=1-\int_{\mathbb{S}\cap\mathbb{A}}d\mu_q= 1-\tau_++\tau_-.
\end{align}
Also required is the quantity $F_{\mathbb{S}\cap \mathbb{A}}[q]$ \eqref{F1} with $r_0 = \omega_-$,
$r_1 = \omega_+$.

\subsection{Proof of Theorem \ref{thm2hw}}
Analogous to the proof of Theorem \ref{thm1}, the complete proof of Theorem \ref{thm2hw} requires consideration of in total nine cases of soft/hard wall conditions, which are given by different combinations of truncated summations. However, since the soft-wall and one-hard-wall cases are just straight forward applications of Corollary \ref{C4.1} together with known results in the literature \cite[Theorem 2.1]{AFLS25} \cite[Theorem 1.1]{BKS23}, we will focus on the two-hard-wall cases only. 
Let $\tau_b=bq'(b)/2,\,\tau_0=q'(1)/2$ be the critical $\tau$-values such that $r_\tau$ lies on the boundaries of the droplet.
As seen in Section 3, the asymptotic behavior of the terms $u_j^{\rm c}$ depends on the range of the index $j$. For $\lfloor N\tau_b\rfloor+1\leq j\leq \lfloor N\tau_0\rfloor $, the critical point lies inside the hard walls; otherwise, it escapes. This implies the separation of the sum
\begin{align*}
    \log {Z}_N^{\rm c}&=\sum_{j=0}^{\lfloor N\tau_b\rfloor}\log u_j^{\rm c}+\sum_{j=\lfloor N\tau_b\rfloor+1}^{\lfloor N\tau_0\rfloor -1}\log u_j^{\rm c}+\sum_{j=\lfloor N\tau_0\rfloor}^{N-1}\log u_j^{\rm c}\\
    &:=S^{\rm out-}+S^{\rm in}+S^{\rm out+} .
\end{align*}
The two-hard-wall cases are then covered by the combinations
\begin{itemize}
    \item Two hard walls at edge:
    \begin{align*}
        S^{\rm in}\bigg\vert_{\substack{\tau_b\to 0\\\tau_0\to 1}}.
    \end{align*}
    \item Left hard wall inside, right hard wall at edge:
    \begin{align*}
        S^{\rm out-}+S^{\rm in}\vert_{\tau_0\to1}.
    \end{align*}
    \item Left hard wall at boundary, right hard wall inside:
    \begin{align*}
        S^{\rm in}\vert_{\tau_b\to0}+S^{\rm out+}.
    \end{align*}
    \item Two hard walls inside:
    \begin{align*}
        S^{\rm out-}+S^{\rm in}+S^{\rm out+}.
    \end{align*}
\end{itemize}
Theorem \ref{thm2hw} is then proved by combining Propositions \ref{Sout-}, \ref{Sin} and \ref{Sout+} below.

\subsection{Computation of $S^{\rm out-}$}
\propo{Sout-}{
The summation $S^{\rm out -}$ admits the following large $N$ expansion
\begin{align*}
    S^{\rm out -}=&N^2(-\tau_bq(b)+\tau_b^2\log b)-\tau_bN\log N-N\tau_b\lrbra{\log(2\tau_b)-1}-b\sqrt{\Delta q(b)}\sqrt{N}\gamma^{\rm out}\\
&-\tilde{\alpha}^{\rm out}-\frac{b\partial_r\Delta q(b)}{\Delta q(b)}\beta^{\rm out}+\frac{b^2\Delta q(b)}{4\tau_b}-\frac{1}{4}\log \pi\\
&+\eta_N^b\left(N(q(b)-2\tau_b\log b)-\frac{1}{2}\log \frac{\pi}{2N\Delta q(b)}+(1+\eta_N^b)\log b\right)+{\rm o}(1),
\end{align*}
where the positive constants $\gamma^{\rm out },\,\tilde{\alpha}^{\rm out},\,\beta^{\rm out} $ are given by \eqref{constout} and $\eta_N^b=\{N\tau_b\}$.
}
\begin{proof}
    According to the large $N$ asymptotics for $u_j^{\rm c}$, we divide $S^{\rm out -}$ into five sums
    \begin{align*}
        S^{\rm out -}=S^{\rm out -}_{0}+S^{\rm out -}_{1-}+S^{\rm out -}_{2-}+S^{\rm out -}_{1+}+S^{\rm out -}_{2+},
    \end{align*}
    where
    \begin{align}\label{Sout--split}
        \begin{aligned}
            &S^{\rm out -}_{0}=\sum_{j=0}^{\lfloor N\tau_b\rfloor-1}-NV_\tau(b)-\log N,\quad S^{\rm out -}_{1-}=-\sum_{j=0}^{\lfloor N(\tau_b-\Delta_N)\rfloor-1}\log V'_\tau(b),\\
        & S^{\rm out -}_{2-}=\sum_{j=0}^{\lfloor N(\tau_b-\Delta_N)\rfloor-1}\log \lrbra{1-\frac{V_\tau''(b)}{NV_\tau'(b)^2}},\quad S^{\rm out -}_{1+}=\sum_{j=\lfloor N(\tau_b-\Delta_N)\rfloor}^{\lfloor N\tau_b\rfloor-1} \log\lrbra{\sqrt{\frac{2N}{\Delta q(b)}}f_1(x_b) },\quad \\
        &S^{\rm out -}_{2+}=\sum_{j=\lfloor N(\tau_b-\Delta_N)\rfloor}^{\lfloor N\tau_b\rfloor-1}\log\lrbra{1-\frac{1}{\sqrt{N}b}\sqrt{\frac{2}{\Delta q(b)}}\frac{h_1^{\rm c}(x_b;b)}{f_1(x)}}.
        \end{aligned}
    \end{align}
    The first sum can be evaluated directly, which gives
    \begin{align}\label{S0out-}
        S^{\rm out -}_{0}=-(N\tau_b-\eta_N^b)(Nq(b)+\log N-N\tau_b\log b+(1+\eta_N^b)\log b).
    \end{align}
    For the second sum, we apply the Euler-Maclaurin formula \eqref{EM1} to get
    \begin{align*}
        \sum_{j=0}^{\lfloor N(\tau_b-\Delta_N)\rfloor-1}\log &V'_\tau(b)=N\int_{0}^{\tau_b-\Delta_N-\xi_N^b/N}\log \lrbra{\frac{2}{b}(\tau_b-\tau)}d\tau \\
        &+\frac{1}{2}\log \lrbra{\frac{2}{b}\tau_b}-\frac{1}{2}\log \lrbra{\frac{2}{b}(\Delta_N+\xi_N^b/N)}+{\rm o}(1)
    \end{align*}
    Calculating the integral explicitly leads to the $N\to \infty $ expansion
    \begin{align}\label{S1-out-}
        \begin{aligned}
            S_{1-}^{\rm out-}=&-(N(\tau_b-\Delta_N)-\xi_N^b)\log \frac{2}{b}-N(\tau_b(\log \tau_b-1)-\Delta_N(\log \Delta_N-1))\\
        &-\frac{1}{2}\log \tau_b+\lrbra{\xi_N^b+\frac{1}{2}}\log \Delta_N+{\rm o}(1).
        \end{aligned}
    \end{align}
    For $S_{2-}^{\rm out-} $, we first expand the logarithm to get
    \begin{align*}
        \sum_{j=0}^{\lfloor N(\tau_b-\Delta_N)\rfloor-1}\log \lrbra{1-\frac{V_\tau''(b)}{NV_\tau'(b)^2}}=-\frac{1}{N}\sum_{j=0}^{\lfloor N(\tau_b-\Delta_N)\rfloor-1}\frac{V_\tau''(b)}{V_\tau'(b)^2}+{\rm o}(1).
    \end{align*}
    The Euler-Maclaurin formula can then be applied, and in view of \eqref{deriV}, we have
    \begin{align}\label{S2-out-}
        S_{2-}^{\rm out-}&=-\int_{0}^{\tau_b-\Delta_N-\xi_N^b/N}\frac{b^2\Delta q(b)+2(\tau-\tau_b)}{4(\tau_b-\tau)^2}d\tau+{\rm o}(1) \nonumber \\
        &=\frac{b^2\Delta q(b)}{4}\lrbra{\frac{1}{\tau_b}-\frac{1}{\Delta_N}}+\frac{1}{2}(\log \tau_b-\log \Delta_N)+{\rm o}(1).
    \end{align}
The summation $S_{1+}^{\rm out-} $ is evaluated in two parts
\begin{align*}
    S_{1+}^{\rm out-}=\sum_{j=\lfloor N(\tau_b-\Delta_N)\rfloor}^{\lfloor N\tau_b\rfloor-1} \log\lrbra{\sqrt{\frac{2N}{\Delta q(b)}} }+\sum_{j=\lfloor N(\tau_b-\Delta_N)\rfloor}^{\lfloor N\tau_b\rfloor-1} \log\lrbra{f_1(x_b) },
\end{align*}
where the first one gives
\begin{align}
    \sum_{j=\lfloor N(\tau_b-\Delta_N)\rfloor}^{\lfloor N\tau_b\rfloor-1} \log\lrbra{\sqrt{\frac{2N}{\Delta q(b)}} }=(N\Delta_N+\xi_N^b-\eta_N^b-1)\log\sqrt{\frac{2N}{\Delta q(b)}}.
\end{align}
To obtain the $N\to \infty$ expansion of the second sum, we start from
    \begin{align*}
    \sum_{j=\lfloor N(\tau_b-\Delta_N)\rfloor}^{\lfloor N\tau_b\rfloor-1}& \log f_1(x_b)=\int_{N(\tau_b-\Delta_N)-\xi_N^b}^{N\tau_b-\eta_N^b}\log f_1(x_b(j))\,dj\\
    &+\frac{1}{2}\lrbra{\log f_1(x_b(\lfloor N(\tau_b-\Delta_N)\rfloor))-\log f_1(x_b(\lfloor N\tau_b\rfloor))}+{\rm o}(1).
\end{align*}
An evaluation of the last two correction terms gives
\begin{align*}
    \frac{1}{2}&\lrbra{\log f_1\lrbra{\facb{N}\lrbra{\Delta_N+\frac{\xi_N^b}{N}}}+\log f_1\lrbra{\facb{N}\frac{\eta_N^b}{N}}}\\
    &\qquad =-\frac{1}{2}\log \frac{\sqrt{\pi}}{2}-\frac{1}{2}\log \facb{}-\frac{1}{2}\log(2\sqrt{N}\Delta_N)+{\rm o}(1).
\end{align*}
For the integral, we proceed with the change of variable $x=\facb{N}(\tau_b-\tau)$ to get
\begin{align*}
    b\sqrt{\frac{N\Delta q(b)}{2}}&\int_{\facb{}\frac{\eta_N^b}{\sqrt{N}}}^{\facb{N}(\Delta_N+\xi_N^b/N)}\log(f_1(x))\,dx\\
    &=b\sqrt{\frac{N\Delta q(b)}{2}}\int_{0}^{\facb{N}(\Delta_N+\xi_N^b/N)}\log(f_1(x))\,dx-\eta_N^b\log\frac{\sqrt{\pi}}{2}+{\rm o}(1),
\end{align*}
where replacing the lower limit by $0$ results in a correction term.
To extend the upper bound to infinity, we need to add and subtract the term $\log(2x)$ to ensure the convergence of the integral and consider the following two integrals
\begin{align}\label{sllogf1}
    b\sqrt{\frac{N\Delta q(b)}{2}}\lrbra{\int_{0}^{\facb{N}(\Delta_N+\xi_N^b/N)}\log(2xf_1(x))\,dx-\int_{0}^{\facb{N}(\Delta_N+\xi_N^b/N)}\log(2x)\,dx},
\end{align}
The first integral can now be extended to $\infty$ and the error is estimated via the expansion 
\begin{align*}
    \log(2x f_1(x))=\log(\sqrt{\pi}xe^{x^2}\erfc(x))\underset{x\to\infty}{=}-\frac{1}{2x^2}+{\rm O}(x^{-4}),
\end{align*}
which gives
\begin{align*}
   b\sqrt{\frac{N\Delta q(b)}{2}}&\int_{0}^{\facb{N}(\Delta_N+\xi_N^b/N)}\log(2xf_1(x))\,dx\\
   &=b\sqrt{\frac{N\Delta q(b)}{2}}\lrbra{\int_0^\infty-\int_{\facb{N}(\Delta_N+\xi_N^b/N)}^\infty} \log(2xf_1(x))\,dx\\
   &=b\sqrt{\frac{N\Delta q(b)}{2}}\int_0^\infty \log(2xf_1(x))\,dx +\frac{b^2\Delta q(b)}{4\Delta_N}+{\rm o}(1).
\end{align*}
The second integral in \eqref{sllogf1} can be explicitly evaluated and gives the asymptotic form
\begin{align*}
    b\sqrt{\frac{N\Delta q(b)}{2}}&\int_{0}^{\frac{1}{b}\sqrt{\frac{2N}{\Delta q(b)}}(\Delta_N+\xi_N^b/N)}\log (2x)\, dx\\
    &=\frac{1}{2}\,(\xi_N^b + N \Delta_{N})\,
\log\!\left(\frac{8N \Delta_{N}^{2}}{b^{2}\,\Delta q(b)}\right)
- N \Delta_{N}
+{\rm o}(1).
\end{align*}
Combining the formulas together, we have
\begin{align*}
    \sum_{j=\lfloor N(\tau_b-\Delta_N)\rfloor}^{\lfloor N\tau_b\rfloor}& \log f_1(x_b)\underset{N\to\infty}{=}-b\sqrt{\Delta q(b)}\sqrt{N}\gamma^{\rm out} +\frac{b^2\Delta q(b)}{4\Delta_N}\\
    &-\lrbra{\xi_N^b+N\Delta_N+\frac{1}{2}}\log\lrbra{\frac{2}{b}\sqrt{\frac{2N}{\Delta q(b)}}\Delta_N}+N\Delta_N-\lrbra{\eta_N^b-\frac{1}{2}}\log\frac{\sqrt{\pi}}{2}+{\rm o}(1),
\end{align*}
where $\gamma^{\rm out}$ is given by \eqref{constout}. Therefore, we have
\begin{align}\label{S1+out-}
    \begin{aligned}
        S_{1+}^{\rm out-}=&-b\sqrt{\Delta q(b)}\sqrt{N}\gamma^{\rm out} +\frac{b^2\Delta q(b)}{4\Delta_N}-(\xi_N^b+N\Delta_N)\log\lrbra{\frac{2}{b}\Delta_N}+N\Delta_N+\xi_N^b\\
    &+\frac{1}{2}\log\frac{2N}{\Delta q(b)}-\frac{1}{2}\log\lrbra{ \frac{4}{b}\sqrt{\frac{2N}{\pi\Delta q(b)}}\Delta_N}-\eta_N^b\log \sqrt{\frac{\pi N}{2\Delta q(b)}}+{\rm o}(1).
    \end{aligned}
\end{align}
As for $S^{\rm out -}_{2+}$, expanding the logarithm and using the Euler-Maclaurin formula yields
\begin{align*}
    \sum_{j=\lfloor N(\tau_b-\Delta_N)\rfloor}^{\lfloor N\tau_b\rfloor-1}\log&\lrbra{1-\frac{1}{\sqrt{N}b}\sqrt{\frac{2}{\Delta q(b)}}\frac{h_1^{\rm c}(x_b;b)}{f_1(x_b)}}\\
    &=-\int_{N(\tau_b-\Delta_N)-\xi_N^b}^{N\tau_b-\eta_N^b}\frac{1}{\sqrt{N}b}\sqrt{\frac{2}{\Delta q(b)}}\frac{h_1^{\rm c}(x(j);b)}{f_1(x(j))}dj+{\rm o}(1).
\end{align*}
After a change of variable $x=\frac{1}{\sqrt{N}b}\sqrt{\frac{2}{\Delta q(b)}}(\tau_b-j/N)$, the integral becomes
\begin{align}
    -&\int_{0}^{\frac{1}{b}\sqrt{\frac{2N}{\Delta q(b)}}\Delta_N}\frac{h_1^{\rm c}(x;b)}{f_1(x)}dx+{\rm o}(1)=-\frac{1}{6}\int_{0}^{\frac{1}{b}\sqrt{\frac{2N}{\Delta q(b)}}\Delta_N}\lrbra{\frac{2x^2-1}{f_1(x)}-4x^3}dx \nonumber \\
    &-\frac{1}{6}\frac{b\partial_r\Delta q(b)}{\Delta q(b)}\int_{0}^{\frac{1}{b}\sqrt{\frac{2N}{\Delta q(b)}}\Delta_N}\lrbra{\frac{x^2+1}{f_1(x)}-x(2x^2+3)}dx+{\rm o}(1).
\end{align}
Here, to ensure convergence when extending the upper limit of integration to $\infty$, we need to add and subtract a term $\frac{3}{x+1}$ in the first integral. This then gives
\begin{align*}
    -\int_{0}^{\frac{1}{b}\sqrt{\frac{2N}{\Delta q(b)}}\Delta_N}&\frac{h_1^{\rm c}(x;b)}{f_1(x)}dx=-\tilde{\alpha}^{\rm out}-\frac{b\partial_r\Delta q(b)}{\Delta q(b)}\beta^{\rm out}+\frac{1}{2}\int_{0}^{\frac{1}{b}\sqrt{\frac{2N}{\Delta q(b)}}\Delta_N}\frac{dx}{x+1}+{\rm o}(1),
\end{align*}
where the constants $\tilde{\alpha}^{\rm out},\, \beta^{\rm out} $ are defined in \eqref{constout}.
Therefore
\begin{align}\label{S2+out-}
    S^{\rm out -}_{2+}=-\tilde{\alpha}^{\rm out}-\frac{b\partial_r\Delta q(b)}{\Delta q(b)}\beta^{\rm out}+\frac{1}{2}\log \lrbra{\frac{1}{b}\sqrt{\frac{2N}{\Delta q(b)}}\Delta_N}+{\rm o}(1).
\end{align}
The proof for Proposition \ref{Sout-} is then completed by adding up \eqref{S0out-} \eqref{S1-out-} \eqref{S1+out-} \eqref{S2-out-} and \eqref{S2+out-}.
\end{proof}

\subsection{Computation of $S^{\rm out+} $}
\propo{Sout+}{
The summation $S^{\rm out +}$ admits the following large $N$ expansion
\begin{align}
    \begin{aligned}
        S^{\rm out +}=&-(1-\tau_0)N(Nq(1)+\log N)-N(1-\tau_0)(\log 2(1-\tau_0)-1)\\
    &-\sqrt{N}\gamma^{\rm out}\sqrt{\Delta q(1)}+\tilde{\alpha}^{\rm out}+\frac{\partial_r\Delta q(1)}{\Delta q(1)}\beta^{\rm out}+\frac{\Delta q(1)}{4(1-\tau_0)}+\frac{1}{4}\log \pi\\
    &+\eta_N^1\lrbra{\frac{1}{2}\log \frac{2N}{\Delta q(1)}+\log \frac{\sqrt{\pi}}{2}-(Nq(1)+\log N)}+{\rm o}(1),
    \end{aligned}
\end{align}
where the positive constants $\gamma^{\rm out },\,\tilde{\alpha}^{\rm out},\,\beta^{\rm out} $ are given by \eqref{constout} and $\eta_N^1=\{N\tau_1\}$.
}

\begin{proof}
    A similar method to that employed in the proof of Proposition \ref{Sout-} applies here, and we omit the proof for brevity.
\end{proof}

\subsection{Computation of $S^{\rm in}$}
\propo{Sin}{
The summation $S^{\rm in}$ admits the following large $N$ expansion
\begin{align}
    \begin{aligned}
        S^{\rm in}&=-N^2 I_{\mathbb{S}\cap \mathbb{A}}^{\rm bulk}[\mu_{q}^{\Omega}]-\frac{1}{2}(1-\eta)N\log N+\frac{N}{2}\bigg(-E_{\mathbb{S}\cap \mathbb{A}}^{\rm bulk}[\mu_{q}^\Omega]-\log 2\, \mathcal T_{1,b}^{-}(xq'(x))+\tm(V_{\tau_x}(x))\\ 
    &+(1-\eta)\log (2\pi)\bigg)
    -\sqrt{N}\tp\lrbra{x\sqrt{\Delta q(x)}}\gamma^{\rm in}+\frac{1}{4}\tm\bigg(\log \frac{2\Delta q(x)}{\pi}\bigg)-\frac{1}{6}\log b\\
    &+\tm\bigg(\frac{x\partial_r\Delta q(x)}{\Delta q(x)}\bigg)\beta^{\rm in}+F_{\mathbb{S}\cap \mathbb{A}}[q]-\log b(\eta_N^b+1)\eta_N^b\\
    &+\tm\lrbra{\eta_N^x\lrbra{NV_{\tau_x}(x)+\frac{1}{2}\log \frac{2N\Delta q(x)}{\pi}}}+{\rm o}(1),
    \end{aligned}
\end{align}
where $\tp,\tm$ and $\eta$ are defined in \eqref{tpm} and \eqref{eta} respectively.
}

\begin{proof}
    It is natural to first divide $S^{\rm in}$ into eight parts
    \begin{align*}
        S^{\rm in}=S^{\rm in0}_{1}+S^{\rm in0}_{2}+S^{\rm in0}_{3}+S^{\rm in0}_{4}+S^{\rm in-}_{1}+S^{\rm in-}_{2}+S^{\rm in+}_{1}+S^{\rm in+}_{2},
    \end{align*}
    where
    \begin{align}\label{SinSplit}
        \begin{aligned}
            &S^{\rm in0}_{1}=\frac{1}{2}\sum_{j=\lfloor N\tau_b\rfloor}^{\lfloor N\tau_0\rfloor-1}\log\frac{2\pi}{N},\quad  S^{\rm in0}_{2}=-\frac{1}{2}\sum_{j=\lfloor N\tau_b\rfloor}^{\lfloor N\tau_0\rfloor-1}\log V_\tau''(r_\tau),\quad
        S^{\rm in0}_{3}=-N\sum_{j=\lfloor N\tau_b\rfloor}^{\lfloor N\tau_0\rfloor-1}V_\tau(r_\tau),\\
        &S^{\rm in0}_{4}=\sum_{j=\lfloor N(\tau_b+\delta_N)\rfloor}^{\lfloor N(\tau_0-\delta_N)\rfloor}\log\lrbra{1+\frac{1}{N}\mathcal{B}^{\rm c}(r_\tau)},\quad
        S^{\rm in-}_{1}=\sum_{j=\lfloor N\tau_b\rfloor}^{\lfloor N(\tau_b+\delta_N)\rfloor-1}\log f_2(x_b),\quad 
        \\ 
        &S^{\rm in-}_{2}=\sum_{j=\lfloor N\tau_b\rfloor}^{\lfloor N(\tau_b+\delta_N)\rfloor-1}\log\lrbra{1-\frac{1}{\sqrt{N}b}\sqrt{\frac{2}{\Delta q(b)}}\frac{h_2^{\rm c}(x_b)}{f_2(x_b)}},\quad 
        S^{\rm in+}_{1}=\sum_{j=\lfloor N(\tau_0-\delta_N)\rfloor+1}^{\lfloor N\tau_0\rfloor-1}\log f_2(x_1),\\
        &S^{\rm in+}_{2}=\sum_{j=\lfloor N(\tau_0-\delta_N)\rfloor+1}^{\lfloor N\tau_0\rfloor-1}\log\lrbra{1+\frac{1}{\sqrt{N}}\sqrt{\frac{2}{\Delta q(1)}}\frac{h_2^{\rm c}(x_1)}{f_2(x_1)}}.
        \end{aligned}
    \end{align}
    The Euler-Maclaurin formula can then be applied to determine the asymptotic expansion for each sum. A direct calculation gives
    \begin{align}\label{Sin01}
        S^{\rm in0}_{1}=\frac{1}{2}(\log 2\pi-\log N)(N(\tau_0-\tau_b)-\eta_N^1+\eta_N^b).
    \end{align}
    Application of the Euler-Maclaurin formula \eqref{EM1} yields
    \begin{align}\label{Sin02}
        \begin{aligned}
            S^{\rm in0}_{2}&=-\frac{N}{2}\int_{\tau_b-\eta_N^b/N}^{\tau_0-\eta_N^1/N}\log(\Delta q(r_\tau))d\tau+\frac{1}{4}\lrbra{\log\Delta q(r_{\tau_0})-\log \Delta q(r_{\tau_b})}+{\rm o}(1)\\
        &=-\frac{N}{2}\lrbra{\frac{1}{2}\int_{b}^{1}r\Delta q(r)\log\lrbra{\Delta q(r)}dr-\frac{1}{N}\tm(\eta_N^x\log \Delta q(x))}\\
        &\qquad +\frac{1}{4}\tm(\log \Delta q(x))+{\rm o}(1)\\
        &=-\frac{N}{2}E_{\mathbb{S}\cap \mathbb{A}}[\mu_Q]+\frac{1}{2}\tm(\eta_N^x\log \Delta q(x))+\frac{1}{4}\tm(\log \Delta q(x))+{\rm o}(1).
        \end{aligned}
    \end{align}
    For $S^{\rm in0}_{1} $, we need to apply the Euler-Maclaurin formula to the second correction term, which leads to
    \begin{align*}
        \begin{aligned}
            \sum_{j=\lfloor N\tau_b\rfloor}^{\lfloor N\tau_0\rfloor-1}V_\tau(r_\tau)=&N\int_{\tau_b-\eta_N^b/N}^{\tau_0-\eta_N^1/N}(q(r_\tau)-2\tau\log r_\tau)d\tau-\frac{1}{2}\bigg(V_\tau(r_\tau)\bigg|_{\tau=\tau_0-\eta_N^1/N}-V_\tau(r_\tau)\bigg|_{\tau=\tau_b-\eta_N^b/N}\bigg)\\
        &+\frac{1}{12N}\bigg(\partial_\tau V_\tau(r_\tau)\bigg|_{\tau=\tau_0-\eta_N^1/N}-\partial_\tau V_\tau(r_\tau)\bigg|_{\tau=\tau_b-\eta_N^b/N}\bigg)+{\rm o}(1).
        \end{aligned}
    \end{align*}
    The Taylor expansion of $V_\tau(r_\tau)$ reads
    \begin{align}\label{TExpV}
        V_\tau(r_\tau)=V_{\tau'}(r_{\tau'})-2\log r_{\tau'}(\tau-\tau')-\frac{2}{r_{\tau'}^2\Delta q(r_{\tau'})}(\tau-\tau')^2+{\rm O}((\tau-\tau')^3), \quad \text{as }\tau\to \tau'.
    \end{align}
    The correction terms are then given by 
    \begin{align*}
        \frac{1}{12N}\bigg(\partial_\tau V_\tau(r_\tau)\bigg|_{\tau=\tau_0-\eta_N^1/N}-\partial_\tau V_\tau(r_\tau)\bigg|_{\tau=\tau_b-\eta_N^b/N}\bigg)=\frac{1}{12N}(-2\log r_{\tau})\bigg|_{\tau_b}^{\tau_0}=\frac{1}{6N}\log b
    \end{align*}
    and 
    \begin{align*}
        \frac{1}{2}\bigg(V_\tau(r_\tau)\bigg|_{\tau=\tau_0-\eta_N^1/N}-V_\tau(r_\tau)\bigg|_{\tau=\tau_b-\eta_N^b/N}\bigg)=\frac{1}{2}\mathcal{T}^-_{1,b}\lrbra{V_{\tau_x}(x)+\frac{2\eta_N^x\log x}{N}}.
    \end{align*}
    The remaining integral involves two parts that can be treated separately. The first of these $\int q(r)d\tau=\frac{1}{2}\int_a^b q(r)r\Delta q(r)\,dr$ can be identified as $I_{\mathbb{S}\cap \mathbb{A}}[\mu_Q]$ up to some other constants, whereas the second requires an integration by parts
    \begin{align*}
        \int_{\tau_a}^{\tau_b}2\tau\log r_{\tau} \,d\tau&=\frac{1}{4}\int_a^b \log r ((rq'(r))^2)'dr=\frac{1}{4}(rq'(r))^2\log r \bigg|_a^b-\frac{1}{4}\int_a^brq'(r)^2\,dr\\
        & =\frac{1}{4}(rq'(r))^2\log r \bigg|_a^b-\frac{1}{4}q(r)rq'(r)\bigg|_{a}^b+\frac{1}{4}\int_a^b q(r)r\Delta q(r)\,dr.
    \end{align*}
    Combining these together, we get
    \begin{align*}
        \int_a^b V_\tau(r_\tau)d\tau=\frac{1}{4}\int_a^bq(r)r\Delta q(r) \,dr+\frac{1}{2}\tau V_\tau(r_\tau)\bigg|_{\tau(a)}^{\tau(b)}.
    \end{align*}
    Therefore, by setting $a=r_{\tau_b-\eta_N^b/N},\, b=r_{\tau_0-\eta_N^1/N}$ and using the expansion \eqref{TExpV}, the large $N$ asymptotic expansion of $\int_{\tau_b-\eta_N^b/N}^{\tau_0-\eta_N^1/N}(q(r_\tau)-2\tau\log r_\tau)d\tau$ is given as
    \begin{align*}
    N&\lrbra{\frac{1}{4}\int_{b}^{1}q(r)r\Delta q(r)\,dr+\frac{1}{2}\mathcal{T}_{1,b}^-(\tau_xV_{\tau_x}(x))}-\tm\lrbra{\eta_N^xV_{\tau_x}(x)}+\frac{2}{N}(\eta_N^b)^2\log b+{\rm o}(1)\\
    &=NI_{\mathbb{S}\cap\mathbb{A}}[\mu_Q]-\tm\lrbra{\eta_N^xV_{\tau_x}(x)}+\frac{1}{N}(\eta_N^b)^2\log b+{\rm o}(1).
    \end{align*}
    Adding the correction terms then gives the asymptotic expansion for $S_3^{\rm in0}$
    \begin{align}\label{Sin03}
        S_3^{\rm in0}=-N^2I_{\mathbb{S}\cap\mathbb{A}}[\mu_Q]+N\lrbra{\tm\lrbra{\eta_N^xV_{\tau_x}(x)}+\frac{1}{2}\tm V_{\tau_x}(x)}-\frac{1}{6}\log b- \eta_N^b(\eta_N^b+1)\log b+{\rm o}(1).
    \end{align}
    Now we turn to $S_4^{\rm in0}$. We first notice that the summand is of order $1/N$, which enables us to extend the summation to $\sum_{j=\lfloor N\tau_b\rfloor}^{\lfloor N\tau_0\rfloor}$ with an ${\rm o}(1)$ error. After expanding the summand, this is equivalent to evaluating the sum $\frac{1}{N}\sum_{j=\lfloor N\tau_b\rfloor}^{\lfloor N\tau_0\rfloor}\mathcal{B}^{\rm c}(r_\tau)$, which for the $N\to \infty $ expansion is approximated by the integral $\int_{\tau_b}^{\tau_0}\mathcal{B}^{\rm c}(r_{\tau})\,d\tau $ up to a constant term. Recall that $\mathcal{B}^{\rm c}(r_{\tau})$ \eqref{Bc} is given by a sum of four terms, with evaluation of the integrals given by
    \begin{align*}
        &\int_{\tau_b}^{\tau_0}\frac{d\tau}{r_{\tau}^2\Delta q(r_\tau)}=\int_b^1\frac{dr}{2r}=-\frac{1}{2}\log b,\quad \int_{\tau_b}^{\tau_0}\frac{\partial_r\Delta q(r_\tau)d\tau}{r_\tau(\Delta q(r_\tau))^2}=\int_{b}^1\frac{\partial_r\Delta q(r)dr}{2\Delta q(r)}=\frac{1}{2}\log \frac{\Delta q(1)}{\Delta q(b)},\\
        &\int_{\tau_b}^{\tau_0}\frac{\partial_r^2\Delta q(r_\tau)d\tau}{(\Delta q(r_\tau))^2}=\int_b^1\frac{r\partial_r^2\Delta q(r)}{2\Delta q(r)}dr=\frac{r\partial_r\Delta q(r)}{2\Delta q(r)}\bigg|_{b}^1-\int_{b}^1 \lrbra{\frac{\partial_r\Delta q(r)}{2\Delta q(r)}-\frac{r(\partial_r\Delta q(r))^2}{2(\Delta q(r))^2}}dr\\
        &\qquad =\frac{1}{2}\lrbra{\frac{\partial_r\Delta q(1)}{\Delta q(1)}-\frac{b\partial_r\Delta q(b)}{\Delta q(b)}}-\frac{1}{2}\log \frac{\Delta q(1)}{\Delta q(b)}+\frac{1}{2}\int_b^1\frac{r(\partial_r\Delta q(r))^2}{(\Delta q(r))^2}dr,\\
        &\int_{\tau_b}^{\tau_0}\frac{(\partial_r\Delta q(r_\tau))^2}{(\Delta q(r_\tau))^3}d\tau=\frac{1}{2}\int_b^1\frac{r(\partial_r\Delta q(r))^2}{(\Delta q(r))^2}dr.
    \end{align*}
    Therefore, we have
    \begin{align}\label{Sin04}
        S^{\rm in0}_4=\frac{1}{12}\log b-\frac{1}{16}\lrbra{\frac{\partial_r\Delta q(1)}{\Delta q(1)}-\frac{b\partial_r\Delta q(b)}{\Delta q(b)}}-\frac{1}{12}\log \frac{\Delta q(1)}{\Delta q(b)}+\frac{1}{24}\int_b^1r\lrbra{\frac{\partial_r\Delta q(r)}{\Delta q(r)}}^2dr+{\rm o}(1).
    \end{align}
    Applying the Euler-Maclaurin formula to $S^{\rm in-}_{1}$, we get
    \begin{align*}
        S^{\rm in-}_{1}=&N\int_{\tau_b-\eta_N^b/N}^{\tau_b+\delta_N-\tilde{\xi}_N^b/N}\log f_2(x_b)d\tau\\
        &-\frac{1}{2}\lrbra{\log f_2\lrbra{\frac{1}{b}\sqrt{\frac{2N}{\Delta q(b)}}(-\delta_N+\tilde{\xi}_N^b/N)}-\log f_2(\eta_N^b/N)}+{\rm o}(1),
    \end{align*}
    where $f_2$ and $x_b$ are given by \eqref{eq:f2h2} and \eqref{x-ome} respectively. Neglecting the higher order terms, the correction terms can be simplified to $\frac{1}{2}\log 2$. For the integral, a further change of variable $\tau\to x_b$ gives
    \begin{align*}
        N\int_{\tau_b-\eta_N^b/N}^{\tau_b+\delta_N-\tilde{\xi}_N^b/N}\log f_2(x_b)d\tau=b\sqrt{\frac{N\Delta q(b)}{2}}\int_{\facb{N}(-\delta_N+\tilde{\xi}_N^b/N)}^{\facb{N}\eta_N^b/N}\log f_2(x)dx.
    \end{align*}
    By the definition of $\delta_N$, the lower limit tends to $-\infty$ as $N\to \infty$. Since $\log f_2(x)$ decays exponentially as $x\to -\infty$, the lower limit can be extended to $-\infty$ with an exponentially small error. For the upper limit, when replacing it by $0$, due to the value $\log f_2(0)=-\log 2$, we pick up a term $-\eta_N^b\log2$. Therefore, we have
    \begin{align}\label{Sin-1}
        \begin{aligned}
            S^{\rm in-}_{1}&=b\sqrt{\frac{N\Delta q(b)}{2}}\int_{-\infty}^0 \log \lrbra{\frac{1}{2}\erfc(x)}dx-\lrbra{\eta_N^b+\frac{1}{2}}\log 2+{\rm o}(1)\\
        &=-b\sqrt{N\Delta q(b)}\gamma^{\rm in}-\lrbra{\eta_N^b+\frac{1}{2}}\log 2+{\rm o}(1).
        \end{aligned}
    \end{align}
    A similar calculation gives
    \begin{align}\label{Sin+1}
        S^{\rm in+}_1=-\sqrt{N\Delta q(1)}\gamma^{\rm in}+\lrbra{\eta_N^1+\frac{1}{2}}\log 2+{\rm o}(1).
    \end{align}
    Expanding the logarithm and applying the Euler-Maclaurin formula, we have
    \begin{align*}
        S^{\rm in-}_2&=-\frac{1}{\sqrt{N}}\sum_{j=\lfloor N\tau_b\rfloor+1}^{\lfloor N(\tau_b+\delta_N)\rfloor-1}\facb{}\frac{h_2^{\rm c}(x_b;b)}{f_2(x_b)}+{\rm o}(1)\\
        &=\int_{-\facb{N}\delta_N}^{0}\lrbra{\frac{1}{6}(2x^2-1)+\frac{1}{6}(x^2+1)\frac{b\partial_r\Delta q(b)}{\Delta q(b)}}\Phi'(x)dx+{\rm o}(1).
    \end{align*}
    The lower limit can again be extended to $-\infty$ due to the exponential decay of the integrand, and it gives
    \begin{align}\label{Sin-2}
        S^{\rm in-}_2=\tilde\alpha^{\rm in}-\frac{b\partial_r\Delta q(b)}{\Delta q(b)}\beta^{\rm in}+{\rm o}(1).
    \end{align}
    A similar evaluation gives
    \begin{align}\label{Sin+2}
        S^{\rm in+}_2=-\tilde\alpha^{\rm in}+\frac{\partial_r\Delta q(1)}{\Delta q(1)}\beta^{\rm in}+{\rm o}(1).
    \end{align}
    The proposition is then proved by adding up the formulas \eqref{Sin01}, \eqref{Sin02}, \eqref{Sin03}, \eqref{Sin04}, \eqref{Sin-1}, \eqref{Sin-2}, \eqref{Sin+1} and \eqref{Sin+2}.
\end{proof}

\section{Appendix: $Q^{\rm c}(Y)=\frac{W}{L}Y^2 $ with an annular gap inside the droplet}\label{AppD}

Here we present the calculation of the large $N$ form of the partition function for the cylinder Coulomb gas with the Gaussian weight $Q^{\rm c}(Y)=WY^2/L$, in the presence of an inner gap. Thus we consider the case that no charge lies in the fixed interval
$$
\left(Y_-,Y_+\right)\subset\left(-\frac{L}{2W},\frac{L}{2W}\right).
$$
Under the exponential change of variables $r=e^{2\pi Y}$, the forbidden interval is the annular gap
$e^{2\pi Y_-}<r<e^{2\pi Y_+}$ inside the droplet. Define
\begin{equation}\label{eq:gap-parameters}
\tau_1:=\frac12-\frac{W}{L}Y_+,
\quad
\tau_2:=\frac12-\frac{W}{L}Y_-,
\quad
d:=\tau_2-\tau_1,
\quad
\tau_*:=\frac{\tau_1+\tau_2}{2},
\quad
\lambda:=\sqrt{\frac{2\pi L}{W}}.
\end{equation}
Thus $0<\tau_1<\tau_2<1$.  We use the normalisations
$$
\vartheta(z\mid\tau):=\sum_{k\in\mathbb Z}e^{2\pi i k z+\pi i k^2\tau},
\qquad
\eta_{\rm D}(\tau):=e^{\pi i\tau/12}\prod_{k=1}^{\infty}(1-e^{2\pi i k\tau})
$$
for the Jacobi theta function and Dedekind eta function respectively. Let $Z_N^{{\rm sc},{\rm gap}}(Q^{\rm c})$ be the corresponding partition function, i.e.
\begin{align}\label{Zscgap}
    Z_N^{{\rm sc},{\rm gap}}(Q^{\rm c})=A_{N,\beta}^{\rm c}  \Big |_{\beta = 2}  W^{2N}  \prod_{l=0}^{N-1} \bigg ( \int_{[-\infty,Y_-]\cup[Y_+,+\infty]} e^{-2 \pi  \rho_{\rm b} W^2 Y^2} e^{2 \pi  (N-2l -1) Y}  \, dY \bigg ),
\end{align}
and $Z_N^{{\rm sc}}(Q^{\rm c})$ be the one given in \eqref{b.4}, then we have the following proposition.

\propo{Pgap}{Let the aspect ratio $L/W$ and the gap endpoints $Y_-,Y_+$ be fixed as $N\to\infty$.  If each $Y_j$ is integrated over
$(-\infty,Y_-]\cup[Y_+,\infty)$, then
\begin{multline}\label{eq:gap-asymptotic}
\log\frac{Z_N^{{\rm sc},{\rm gap}}(Q^{\rm c})}{Z_N^{{\rm sc}}(Q^{\rm c})}
=-\frac{\lambda^2d^3}{12}N^2-\frac d2N\log N
+d\left(1-\log\big(\sqrt\pi\,\lambda d\big)\right)N\\
-\frac{2\sqrt2}{\lambda}(\gamma^{\rm in}+\gamma^{\rm out})\sqrt N
+\frac{2}{\lambda^2d}
+\log\frac{\vartheta\left(N\tau_*\,\middle|\,\frac{\pi i}{\lambda^2d}\right)}
{\eta_{\rm D}\left(\frac{\pi i}{\lambda^2d}\right)}+{\rm o}(1).
\end{multline}
In particular, there is no separate $\log N$ term, while the constant-order term has a bounded oscillatory dependence on $N$ through the theta function.  Upon adding \eqref{b.4},
\begin{multline}\label{eq:gap-full}
\log Z_N^{{\rm sc},{\rm gap}}(Q^{\rm c})
=\frac{\lambda^2}{12}(1-d^3)N^2-\frac{1+d}{2}N\log N\\
+N\left(\log(\sqrt\pi\,\lambda)
+d\left(1-\log(\sqrt\pi\,\lambda d)\right)\right)
-\frac{2\sqrt2}{\lambda}(\gamma^{\rm in}+\gamma^{\rm out})\sqrt N\\
-\frac{\lambda^2}{12}+\frac{2}{\lambda^2d}
+\log\frac{\vartheta\left(N\tau_*\,\middle|\,\frac{\pi i}{\lambda^2d}\right)}
{\eta_{\rm D}\left(\frac{\pi i}{\lambda^2d}\right)}+{\rm o}(1).
\end{multline}
}

\begin{proof}
Let
$$
x_{l,N}:=\frac{l+\frac12}{N},
\qquad l=0,\ldots,N-1.
$$
Completing the square in the $l$-th integral in \eqref{Zscgap}, and using
$\rho_{\rm b}W^2=NW/L$, shows that its Gaussian centre is
$$
m_{l,N}=\frac{N-2l-1}{2\rho_{\rm b}W^2}
=\frac{L}{2W}(1-2x_{l,N}).
$$
After a change of variable $\sqrt{2 \pi  \rho_{\rm b}}W(Y-m_{l,N})\to s$ in the integral in \eqref{Zscgap} and division by the unrestricted Gaussian integral, the remaining term is therefore
\begin{equation}\label{eq:gap-product}
\frac{Z_N^{{\rm sc},{\rm gap}}(Q^{\rm c})}{Z_N^{{\rm sc}}(Q^{\rm c})}
=\prod_{l=0}^{N-1}p_{l,N},
\qquad
p_{l,N}
=\frac12\erfc\left(\lambda\sqrt N(x_{l,N}-\tau_1)\right)
+\frac12\erfc\left(\lambda\sqrt N(\tau_2-x_{l,N})\right). 
\end{equation}
For $x_{l,N}<\tau_*$ the first term in $p_{l,N}$ is dominant, whereas for
$x_{l,N}>\tau_*$ the second is dominant.  The modes nearest $\tau_*$ require both terms and will produce the theta factor.
Denote
$$
S_N:=\log\frac{Z_N^{{\rm sc},{\rm gap}}(Q^{\rm c})}{Z_N^{{\rm sc}}(Q^{\rm c})}
=\sum_{l=0}^{N-1}\log p_{l,N}.
$$
Accordingly, select the first tail in \eqref{eq:gap-product} for $l\le m$ and the second for $l\ge m+1$, and decompose exactly
$$
S_N=B_N+C_N,
$$
where
\begin{align}
    \begin{aligned}
B_N:={}&\sum_{l=0}^{m}
\log\left[\frac12\erfc\left(\lambda\sqrt N(x_{l,N}-\tau_1)\right)\right]
+\sum_{l=m+1}^{N-1}
\log\left[\frac12\erfc\left(\lambda\sqrt N(\tau_2-x_{l,N})\right)\right],\\
C_N:={}&\sum_{l=0}^{m}
\log\left(1+
\frac{\erfc\left(\lambda\sqrt N(\tau_2-x_{l,N})\right)}
{\erfc\left(\lambda\sqrt N(x_{l,N}-\tau_1)\right)}\right)+\sum_{l=m+1}^{N-1}
\log\left(1+
\frac{\erfc\left(\lambda\sqrt N(x_{l,N}-\tau_1)\right)}
{\erfc\left(\lambda\sqrt N(\tau_2-x_{l,N})\right)}\right).
\end{aligned}
\end{align}
Let
$$
j_*:=N\tau_*-\frac12,
\quad
m:=\lfloor j_*\rfloor,
\quad
\theta_N:=\{j_*\},
\quad
h:=N^{-1/2},
\quad
D:=\frac d2,
\quad
A:=D\sqrt N,
$$
and make the change of summation variable
\begin{enumerate}
    \item If $l\leq m$: $l=m-k$ with $0\le k\le m$
    $$
    \sqrt N(x_{l,N}-\tau_1)=A-h(k+\theta_N),
    \qquad
    \sqrt N(\tau_2-x_{l,N})=A+h(k+\theta_N).
    $$
    \item If $l\geq m+1$: $l=m+1+k$ with $0\le k\le N-m-2$
    $$
    \sqrt N(\tau_2-x_{l,N})=A-h(k+1-\theta_N),
    \qquad
    \sqrt N(x_{l,N}-\tau_1)=A+h(k+1-\theta_N).
    $$
\end{enumerate}
Then the first sum $B_N$ can be shown to satisfy
$$
B_N=T_{\theta_N}(h,A)+T_{1-\theta_N}(h,A)+{\rm O}(e^{-cN}),
$$
where
\begin{align}
    T_\alpha(h,A):=\sum_{k=0}^{\infty}g(A-h(k+\alpha)),\quad g(z):=\log\left(\frac12\erfc(\lambda z)\right).
\end{align}
Here extending the sums to infinity only adds values of $g(z)$ with $z\le-c\sqrt N$ for some constant $c$. Since $g(z)={\rm O}(e^{-\lambda^2z^2}/|z|)$ as $z\to-\infty$, and since both gap endpoints have a fixed positive distance from the soft droplet endpoints, the total extension error is ${\rm O}(e^{-cN})$.
Write $I(A):=\int_{-\infty}^{A}g(z)\,dz$.  The Euler-Maclaurin summation, carried through the next Bernoulli term, gives uniformly for $\alpha\in[0,1]$,
\begin{equation}\label{eq:gap-EM}
T_\alpha(h,A)=\frac1hI(A)-B_1(\alpha)g(A)+\frac h2B_2(\alpha)g'(A)
-\frac{h^2}{6}B_3(\alpha)g''(A)+R_{\alpha,N},
\end{equation}
where $B_1(\alpha)=\alpha-1/2$, $B_2(\alpha)=\alpha^2-\alpha+1/6$ and $B_3(\alpha)=\alpha^3-3\alpha^2/2+\alpha/2$.  Since $g''(A)={\rm O}(1)$ and $g'''$ is integrable, the last two terms satisfy
$$
\frac{h^2}{6}B_3(\alpha)g''(A)={\rm O}(N^{-1}),
\qquad
|R_{\alpha,N}|\le Ch^2\int_{-\infty}^{A}|g'''(z)|dz={\rm O}(N^{-1}).
$$
They may therefore be absorbed into ${\rm o}(1)$.  Notice that stopping with a remainder involving $\int|g''|$ would not suffice, since $g''(z)\to-2\lambda^2$ as $z\to+\infty$.  Since
$$
B_1(1-\alpha)=-B_1(\alpha),
\qquad B_2(1-\alpha)=B_2(\alpha),
$$
the $g(A)$ terms cancel and
$$
B_N=\frac2hI(A)+hB_2(\theta_N)g'(A)+{\rm o}(1).
$$
Since $A\sim {\rm O}(\sqrt{N})$ as $N\to \infty$, a further expansion can be performed.
The asymptotic expansion of the complementary error function gives
$$
g(z)=-\lambda^2z^2-\log z-\log(2\sqrt\pi\lambda)
-\frac{1}{2\lambda^2z^2}+{\rm O}(z^{-4}),\quad \text{as } z\to+\infty.
$$
Consider the regularised integral
$$
K_\lambda:=\int_{-\infty}^{0}g(z)\,dz
+\int_0^\infty\left(g(z)+\lambda^2z^2+\log z
+\log(2\sqrt\pi\lambda)\right)dz,
$$
which, by the definitions \eqref{gi}, can be written as $-\frac{\sqrt2}{\lambda}(\gamma^{\rm in}+\gamma^{\rm out})$.
The asymptotic expansion for $I(A)$ is given by
$$
I(A)=-\frac{\lambda^2A^3}{3}-A\log A
+(1-\log(2\sqrt\pi\lambda))A+K_\lambda
+\frac{1}{2\lambda^2A}+{\rm O}(A^{-3}).
$$
Using $hg'(A)=-\lambda^2d+o(1)$ and $A=d\sqrt N/2$, we obtain
\begin{multline}\label{eq:gap-B}
B_N=-\frac{\lambda^2d^3}{12}N^2-\frac d2N\log N
+d\left(1-\log(\sqrt\pi\,\lambda d)\right)N
-\frac{2\sqrt2}{\lambda}(\gamma^{\rm in}+\gamma^{\rm out})\sqrt N\\
+\frac{2}{\lambda^2d}-\lambda^2dB_2(\theta_N)+{\rm o}(1).
\end{multline}
Now we turn to the remaining sum $C_N$. For $v\ge0$ define the ratio
\begin{equation}\label{eq:gap-ratio-R}
\mathcal R_N(v):=
\frac{\erfc\left(\lambda\sqrt N(D+v/N)\right)}
{\erfc\left(\lambda\sqrt N(D-v/N)\right)}.
\end{equation}
The preceding identities for the two sides of the switching index give the exact finite-sum representation
\begin{equation}\label{eq:gap-C-finite}
C_N={}\sum_{k=0}^{m}\log\left(1+\mathcal R_N(k+\theta_N)\right)
+\sum_{k=0}^{N-m-2}\log\left(1+\mathcal R_N(k+1-\theta_N)\right).
\end{equation}
Introduce the cutoff scale
$$
K_N:=\left\lfloor(\log N)^2\right\rfloor.
$$
For $0\le v\le K_N$ and $N$ sufficiently large, both $D-v/N$ and $D+v/N$ are bounded below by $D/2$.  Thus the expansion
$$
\frac12\erfc(X)=\frac{e^{-X^2}}{2\sqrt\pi X}
\left(1-\frac{1}{2X^2}+{\rm O}(X^{-4})\right),
\qquad X\to+\infty,
$$
may be applied uniformly to
$$
X_\pm:=\lambda\sqrt N(D\pm v/N).
$$
Since
$$
X_+^2-X_-^2=4\lambda^2Dv=2\lambda^2dv,
$$
we obtain, uniformly for $0\le v\le K_N$,
\begin{equation}\label{eq:gap-Mills-uniform}
\begin{aligned}
\mathcal R_N(v)
&=e^{-2\lambda^2dv}\frac{D-v/N}{D+v/N}\left(1+{\rm O}(N^{-1})\right)\\
&=q^v\left(1+{\rm O}\left(\frac{v+1}{N}\right)\right),
\qquad q:=e^{-2\lambda^2d}\in(0,1).
\end{aligned}
\end{equation}
In particular, the error in \eqref{eq:gap-Mills-uniform} is summable over all switching modes.  Since $x\mapsto\log(1+x)$ is Lipschitz with constant one on $[0,\infty)$,
$$
\sum_{k=0}^{K_N}
\left|\log(1+\mathcal R_N(k+\theta_N))
-\log(1+q^{k+\theta_N})\right|
\le \frac{C}{N}\sum_{k=0}^{K_N}(k+1)q^k
={\rm O}(N^{-1}),
$$
uniformly in $\theta_N\in[0,1)$. Thus the presence of $K_N={\rm O}((\log N)^2)$ terms does not enlarge the total error to order one; the geometric factor $q^k$ makes their accumulated error ${\rm O}(N^{-1})$. 
It remains to justify removal of the cutoff.  For $K_N<v\le ND/2$, the elementary upper and lower Mills inequalities give
$$
0\le\mathcal R_N(v)\le Cq^v.
$$
For $v>ND/2$, the numerator in \eqref{eq:gap-ratio-R} is a Gaussian tail at distance at least $3D\sqrt N/2$.  If the denominator still has a positive argument, the same Mills inequalities show that the ratio is ${\rm O}(e^{-cN})$; if its argument is non-positive, the denominator is at least $1/2$ and the same bound follows directly.  Since each finite sum in \eqref{eq:gap-C-finite} contains at most $N$ terms,
$$
\sum_{v>K_N}\log(1+\mathcal R_N(v))
={\rm O}(q^{K_N})+{\rm O}(Ne^{-cN})
={\rm O}\left(e^{-c(\log N)^2}\right).
$$
We have therefore proved the estimate
\begin{equation}\label{eq:gap-C}
\begin{aligned}
C_N={}&\sum_{k=0}^{\infty}\log(1+q^{k+\theta_N})
+\sum_{k=0}^{\infty}\log(1+q^{k+1-\theta_N})\\
&+{\rm O}(N^{-1})+{\rm O}\left(e^{-c(\log N)^2}\right).
\end{aligned}
\end{equation}
We finally rewrite the two infinite products. With
$$
(z;q)_\infty:=\prod_{k=0}^{\infty}(1-zq^k),
$$
equation \eqref{eq:gap-C} reads
$$
C_N=\log\left((-q^{\theta_N};q)_\infty
(-q^{1-\theta_N};q)_\infty\right)+{\rm o}(1).
$$
Set
$$
a:=\lambda^2d,
\qquad
\tau:=\frac{ia}{\pi},
\qquad
u:=\theta_N-\frac12,
$$
so that $q=e^{-2a}=e^{2\pi i\tau}$.  In the convention used here, the Jacobi triple product formula is
\begin{equation}\label{eq:gap-triple-product}
\vartheta(z\mid\tau)
=(q;q)_\infty
\left(-q^{1/2}e^{2\pi iz};q\right)_\infty
\left(-q^{1/2}e^{-2\pi iz};q\right)_\infty.
\end{equation}
Taking $z=\tau u$ gives
$$
q^{1/2}e^{2\pi iz}=q^{\theta_N},
\qquad
q^{1/2}e^{-2\pi iz}=q^{1-\theta_N},
$$
and hence
\begin{equation}\label{eq:gap-C-theta-1}
C_N=\log\vartheta(\tau u\mid\tau)-\log(q;q)_\infty+{\rm o}(1).
\end{equation}
The eta function enters through the remaining Euler product.  Indeed, its definition gives
$$
\eta_{\rm D}(\tau)
=e^{\pi i\tau/12}\prod_{k=1}^{\infty}(1-e^{2\pi i k\tau})
=q^{1/24}(q;q)_\infty.
$$
Since $q=e^{-2a}$, equation \eqref{eq:gap-C-theta-1} becomes
\begin{equation}\label{eq:gap-C-theta-2}
C_N=\log\vartheta(\tau u\mid\tau)
-\log\eta_{\rm D}(\tau)-\frac{a}{12}+{\rm o}(1).
\end{equation}

We use the modular transformations
$$
\vartheta(z\mid\tau)
=(-i\tau)^{-1/2}e^{-\pi iz^2/\tau}
\vartheta\left(\frac z\tau\,\middle|\,-\frac1\tau\right),
\qquad
\eta_{\rm D}\left(-\frac1\tau\right)
=\sqrt{-i\tau}\,\eta_{\rm D}(\tau).
$$
Since $-1/\tau=\pi i/a$ and $z=\tau u$, the first identity gives
$$
\log\vartheta(\tau u\mid\tau)
=-\frac12\log\frac{a}{\pi}+au^2
+\log\vartheta\left(u\,\middle|\,\frac{\pi i}{a}\right),
$$
whereas the second gives
$$
\eta_{\rm D}\left(\frac{\pi i}{a}\right)
=\sqrt{\frac{a}{\pi}}\,\eta_{\rm D}(\tau).
$$
Finally, $B_2(\theta_N)=u^2-1/12$.  Subtracting $aB_2(\theta_N)$ from \eqref{eq:gap-C-theta-2} cancels both $au^2$ and the two $a/12$ terms, while the square-root prefactor is absorbed by the eta transformation.  Therefore
\begin{equation}\label{eq:gap-C-theta-final}
C_N-aB_2(\theta_N)
=\log\frac{\vartheta\left(u\,\middle|\,\frac{\pi i}{a}\right)}
{\eta_{\rm D}\left(\frac{\pi i}{a}\right)}+{\rm o}(1).
\end{equation}
Since $u\equiv N\tau_*\pmod 1$, combining this with \eqref{eq:gap-B} proves \eqref{eq:gap-asymptotic}.  Equation \eqref{eq:gap-full} follows on adding the exact soft-wall expression \eqref{b.4}.
\end{proof}

\providecommand{\bysame}{\leavevmode\hbox to3em{\hrulefill}\thinspace}
\providecommand{\MR}{\relax\ifhmode\unskip\space\fi MR }
\providecommand{\MRhref}[2]{%
  \href{http://www.ams.org/mathscinet-getitem?mr=#1}{#2}
}
\providecommand{\href}[2]{#2}

\end{document}